# Design of dye-sensitized TiO2 materials for photocatalytic hydrogen production: light and shadow

Lorenzo Zani,[a] Michele Melchionna,[b] Tiziano Montini,[b] Paolo Fornasiero [b,*]

[a] *Institute of Chemistry of Organometallic Compounds (CNR-ICCOM), Sesto Fiorentino 50019, Italy.*

[b] *Department of Chemical and Pharmaceutical Sciences, CNR-ICCOM Trieste Research Unit and INSTM Research Unit, University of Trieste, Trieste 34127, Italy; email: pfornasiero@units.it.*

**Table of Contents**

*Abstract:*

Visible light-driven production of fuels and value-added chemicals is currently one of the most intensely investigated research topics across various scientific disciplines, due to its potential to ease the World's dependence on fossil fuels. In this perspective, we recapitulate some of the main features of dye-sensitized photocatalytic systems aimed at solar $H_2$ production, focusing in particular on $TiO_2$-based three-component assemblies with organic sensitizers. Relevant aspects include the structural and electronic properties of the sensitizers, the nature of the semiconductor and the hydrogen evolution catalysts, the role of the sacrificial donor and the effect of the reaction parameters on $H_2$ production rate and stability. Besides presenting the most significant recent developments of the field, we also analyse some of its common practices in terms of experimental design, laboratory procedures and data presentation, trying to highlight their weaknesses and suggesting possible improvements. We then conclude with a short paragraph discussing the possible future development of this exciting research area.

# 1. Introduction: visible light absorption and the relationship with DSSCs

Due to the urgent need to replace fossil fuels as the World's main energy source, the conversion of solar radiation into chemical energy in the form of so-called "solar fuels", often referred to as "artificial photosynthesis",[1] is currently of utmost scientific and technological relevance.[2] Among the different artificial photocatalytic processes, $H_2$ production through water splitting (WS) has probably been the most intensely studied, since $H_2$ is endowed with high volumetric energy density, no carbon footprint and can be either directly burned or used in fuel cells to produce electricity, thus constituting an almost ideal energy carrier.[3,4]

In 1972, the pioneering work of Honda and Fujishima demonstrated that WS into $H_2$ and $O_2$ could be achieved by irradiating a $TiO_2$ photoanode connected to a platinum cathode in an electrochemical cell.[5] The main drawback of such system was the use of a wide band-gap ($\geq 3.0$ eV) semiconductor (SC) as the light-harvesting material, which hampered absorption and conversion of visible light ($\lambda > 400$ nm) and made it necessary to use UV radiation to drive the reaction forward.

To solve such an issue, several possible approaches to modify inorganic heterogeneous photocatalysts have been investigated,[6] including the use of narrow band-gap semiconductors[7,8] the chemical modification of large band-gap materials to impart them the ability to absorb visible light,[9] or the application of more complex photocatalytic assemblies such as Z-schemes.[10,11] Besides, another effective strategy has been the sensitization of semiconductors with molecular dyes, able to harvest light in the desired wavelength range and inject the resulting photogenerated electrons into the SC conduction band.[12] This concept was first established in Dye-Sensitized Solar Cells (DSSC), in which, after excitation and charge injection, electrons are collected at a $TiO_2$ photoanode while holes are transferred to the reduced form of a suitable redox mediator (typically $I_3^-/I^-$), which is then regenerated at the cathode to close the cycle and produce an electric current (Figure 1a).[13] Due to the analogy with DSSCs, such photocatalytic systems are usually called Dye-Sensitized Photocatalysts (DSP).

DSSC and DSP share the same dye/semiconductor interface but, in the latter, photogenerated electrons in the conduction band are transferred to an electrocatalyst (such as Pt) for solar fuel production, instead of being used for electricity generation. Accordingly, in DSP the oxidized dye molecules must be reduced by a suitable hole scavenger to allow the process to continue (Figure 1b). In a proper WS procedure, electrons would be supplied by water itself, allowing coupling of $H_2$ production with $O_2$ evolution without formation of any other by-product. However, water oxidation demands combining the sensitizers with appropriate catalysts, often a synthetically demanding operation,[14] and is usually affected by significant drawbacks, such as the need for a large overpotential,[15] the quick recombination of photogenerated charge carriers, and the rapid back reaction between $H_2$ and $O_2$.[16] Consequently, to achieve a high yield of $H_2$ production, but also to better determine the photocatalyst intrinsic activity, dye regeneration is more commonly carried out by means of a sacrificial electron donor, usually abbreviated as SED (*see below*).[17] From the above discussion it is clear that, despite the similarities between the two systems, the materials used in DSSC or DSP, such as dyes and semiconductors, must work under different conditions and thus will need to be developed independently to provide optimal performances in each application.

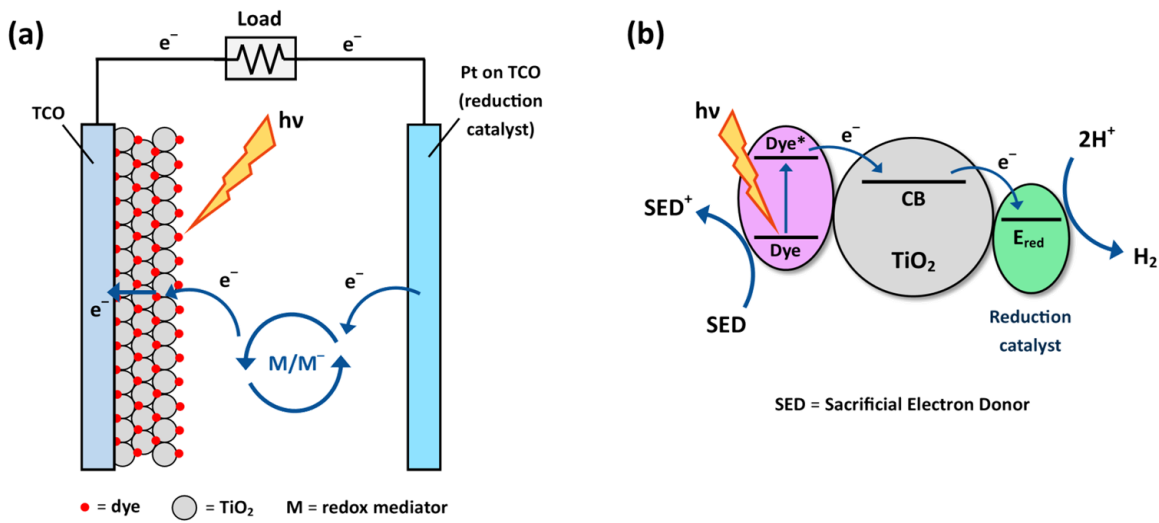

*Figure 1.* *Schematic representation of the working mechanisms of (a) a DSSC and (b) a DSP.*

In this perspective, the main features of selected DSP systems especially developed for $H_2$ production will be critically presented, trying to highlight their strengths as well as the areas where there is still room for improvement. Although the concept of dye-sensitization has been applied to different kinds of materials,[18] inorganic[19] as well as organic,[20,21] we will focus on $TiO_2$-based systems, since they are by far the most investigated in the literature and thus can be more easily compared. In the majority of such systems, $TiO_2$ is used as the anatase crystalline form, or as the commercially available 80:20 anatase/rutile mixture known as P25,[22] but alternatives have also been described.

Before starting the discussion, we will briefly introduce the topic of the correct presentation and comparison of photocatalytic data, as the adoption of a more homogeneous and shared standard will be of crucial importance for the future development of the field.

## 2. How to properly compare data

The criticality of correctly evaluating the merits of a new proposed photocatalyst is one of the contemporary topics of discussions within the photocatalysis community.[23,24] In this respect, the importance of introducing more comprehensive and diligent practices when assessing and comparing photocatalysts performances has been recently highlighted.[25] As an example, for heterogeneous photocatalysts it is very common to report the rate of evolved product per mass of photocatalyst, which gives also an idea of how stable it is, but does not take into account the contribution of its textural features. Hence, adding a rate normalized by the material surface area provides a more thorough screening and is therefore recommended.[26]

Turnover numbers (TON), or in alternative turnover frequency (TOF), are two other classic catalytic parameters, which in the specific case of DSP are calculated over the number of moles of dye covering the semiconductor nanoparticles. Hence, the catalytic sites are assumed to be equal to the number of molecules of dyes, which may in principle lead to underestimated TON or TOF, if not all the dye molecules are in the right spatial configuration for electron transfer to the SC (*e.g.* formation of aggregates, intermolecular charge transfer). Moreover, apart from depending on several conditions such as temperature and pH, TOF is a kinetic-dependent parameter, so that it should be evaluated at low conversions (or at least the reactant or product concentrations should be provided), or ideally as an instantaneous value measured at specified product (or reactant) concentrations.[27] Most studies on heterogeneous photocatalysts, however, compare them in terms of quantum yield (QY), quantum efficiency (QE), or photonic efficiencies, where instead of the number of catalytic sites (which for DSP is assumed to be the number of dye molecules), the number of incident photons is considered.[23] While the adoption of this parameter seems to remove the uncertainty on the effectiveness of the adsorbed dye to transfer electrons, it also introduces an element of ambiguity related to the heterogeneous nature of the photocatalyst. In fact, for heterogeneous systems, not all the incident photons are necessarily absorbed, with scattering phenomena taking place and decreasing the amount of utilized photons. For heterogeneous photocatalysts, the term apparent quantum yield (AQY) is therefore more sensible.[28] As a result, the AQY is almost always an underestimation of the real QY. Moreover, as the AQY is a function of the excitation wavelength,[29] a fact that is at times ignored, the comparison between DSP with different absorption characteristics may be affected by inaccuracies, causing false esteems. A good practice will therefore be to plot the wavelength-dependent AQY profile by measuring it at successively increasing wavelengths, and then verify that the pattern follows that of the photocatalyst or sensitizer absorption spectrum.

## 3. Charge transfer processes in DSP

As shown in Figure 1, the photocatalytic cycle in DSP is initiated by the two steps of light absorption and charge separation, which are of pivotal importance to determine the efficiency of $H_2$ generation. Two main mechanisms have been proposed for the charge separation process involving a photoexcited dye, the semiconductor (most commonly $TiO_2$) and the SED, classified as reductive quenching and oxidative quenching, respectively.[30] Following light absorption (eq. 1), the reductive quenching mechanism proceeds with an electron transfer from the SED to the excited dye, which is thus converted into a radical anion ($D^{\bullet-}$, eq. 2). Subsequent electron injection into the semiconductor conduction band restores the dye in its ground state and completes charge separation (eq. 3). In the oxidative quenching mechanism, on the other hand, the first electron transfer step involves charge injection from the excited dye to the semiconductor, with concomitant formation of a dye radical cation ($D^{\bullet+}$, eq. 4). The latter is then reduced by the SED, and the same charge separation state is reached (eq. 5). Finally, the electrons in the conduction band of the semiconductor will be used for $H_2$ generation by proton reduction (eq. 6). Besides these productive electron transfer events, however, it must be pointed out that detrimental, reverse charge transfer processes can also take place, such as charge recombination between injected electrons and the dye cation or the oxidized SED. The ratio between the rates of forward and backward electron transfer processes is what ultimately dictates the efficiency of the photocatalytic system.[12]

**Photoexcitation:** $\text{Pt/TiO}_2\text{/D} + h\nu \rightarrow \text{Pt/TiO}_2\text{/D*}$  (1)

**Reductive quenching:** $\text{Pt/TiO}_2\text{/D*} + \text{SED} \rightarrow \text{Pt/TiO}_2\text{/D}^{\bullet-} + \text{SED}^+$  (2)

$\text{Pt/TiO}_2\text{/D}^{\bullet-} \rightarrow \text{Pt/TiO}_2(e^-)\text{/D}$  (3)

**Oxidative quenching:** $\text{Pt/TiO}_2\text{/D*} \rightarrow \text{Pt/TiO}_2(e^-)\text{/D}^{\bullet+}$  (4)

$\text{Pt/TiO}_2(e^-)\text{/D}^{\bullet+} + \text{SED} \rightarrow \text{Pt/TiO}_2(e^-)\text{/D} + \text{SED}^+$  (5)

**Proton reduction:** $\text{Pt/TiO}_2(e^-)\text{/D} + H^+ \rightarrow \text{Pt/TiO}_2\text{/D} + 1/2\ H_2$  (6)

In general, the oxidative quenching mechanism is considered to be predominant for almost all dye classes (see below), in analogy with what happens in DSSCs. The reductive quenching mechanism is probably relevant only in the case of poorly reducing, cationic dyes (e. g. thionine, methylene blue, nile blue A), and has been suggested to provide inferior results in terms of $H_2$ production.[30] The efficiency of the charge transfer process between the dye and the semiconductor is usually assessed by means of photoluminescence decay studies: by comparing the excited state lifetime of the dye in solution to that of the dye/semiconductor assembly the rate constant for electron transfer can be calculated with good approximation.[31]

## 4. Design of dye scaffold for application in DSP

Dye design is certainly one of the most important factors affecting light harvesting and charge transfer efficiency in DSP systems. As mentioned in the introduction, the concept of dye-sensitization in DSP was originally derived from that of DSSCs. Consequently, the main classes of dyes employed in photocatalysis resemble those already applied in solar cells, namely (i) metalorganic complexes, especially based on ruthenium with bi- o terpyridine ligands; (ii) porphyrins and phthalocyanines bearing different central metals; (iii) metal-free organic dyes. This latter class of dyes has been the subject of the largest number of studies in recent years,[32,33] and can be further divided into sub-categories, such as emissive dyes traditionally used in chemical biology, or donor (D)-acceptor (A) structures, where electron-donating groups are connected to electron-accepting units *via* conjugated sections of various nature. Such an arrangement allows extending and strengthening the absorption spectra of the resulting compounds, improving their light-harvesting ability. In their simplest form, these compounds are usually denoted as D-π-A dyes, with the electron acceptor also fulfilling the role of anchoring group to the semiconductor;[34] from them, more complex architectures such as D-A-π-A, D-D-π-A and others[35] have been derived by the insertion of additional donor or acceptor units in various parts of the structure.

A comprehensive review on the use of all the above classes of sensitizers in DSP is beyond the scope of this manuscript, and has already been presented elsewhere.[12] Here, we will focus our attention on the employment of organic dyes, since they have provided the best results in DSP systems. Moreover, the fact that they do not contain precious or toxic heavy metals makes them more sustainable than metalorganic complexes, which is particularly relevant in the field of renewable energy technologies. Finally, they are usually accessible through simple and modular synthetic processes, allowing to efficiently tune their stereoelectronic properties. Accordingly, they constitute an ideal platform to analyse how their structural and compositional changes can affect the overall performances of the DSP systems.

Despite their similarities, organic dyes used in DSP have evolved independently from those employed in DSSCs, due to the different conditions in which they must operate. For example, they have to work efficiently in the aqueous environment used in DSP, while DSSC usually contain organic solvents and are moisture-sensitive. In addition, they must be regenerated by SED molecules, which are different from the redox couples commonly used in DSSCs, thus requiring a different alignment of their energy levels (especially the HOMO level). Finally, the fact that DSSCs are self-contained devices in which the photo- and electroactive materials are supported onto electrodes while DSPs operate in heterogeneous suspension bears different requirements in terms of dyes molar absorptivity and loading onto the semiconductor. For the above reasons, although DSP and DSSC dyes usually have similar light harvesting properties, they can present significant differences in the way they are attached to the SC surface, in the balance of their hydrophobic and hydrophilic characteristics, and in their electrochemical properties. Some of these features will be discussed below.

To work efficiently in a DSP system, any sensitizer has to fulfil two obvious requirements: (i) it should be able to absorb light efficiently in the visible region of the spectrum (where solar radiation is maximized); (ii) it should transfer easily the photogenerated electrons to the conduction band of the semiconductor. Thanks to the extensive experience accumulated in the field of DSSC, it has been relatively straightforward to build a library of organic photosensitizers for DSP applications having both these properties.[12,32] By changing the nature of the main chromophore and adjusting the length of the conjugated section, compounds have been reported with the main absorption band going from around 400 nm (as in the case of simple D-A structures, Figure 2, **1**)[36,37] to almost 700 nm, close to the near-IR region (as in the case of porphyrin- or BODIPY-containing species, Figure 2, **2**).[31,38,39] On the other hand, insertion in the structure of additional electron-donating or accepting fragments can serve to modulate the frontier orbitals energy levels and thus alter the rate of the intermolecular charge transfer processes.[40,41] It should be noted that all such properties can be efficiently modelled by means of DFT calculations.[42]

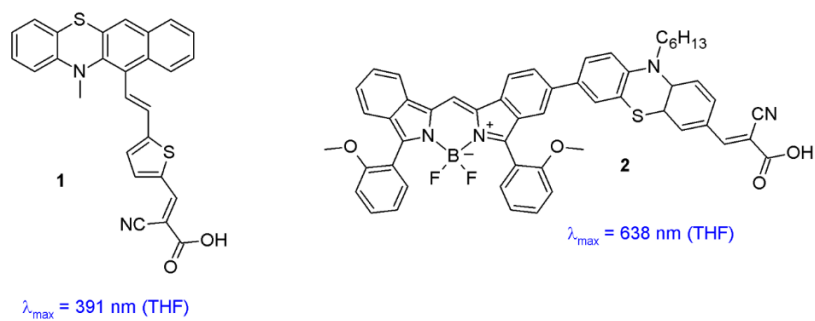

*Figure 2. Structures of organic dyes having very different UV-Vis absorption maxima in solution.*

Nevertheless, there are other characteristics the sensitizers must possess to boost the activity of a three-component photocatalytic system. First, to be able to transfer electrons quickly to the semiconductor, they should have a robust anchoring to its surface. Due to the analogy with DSSC dyes, most compounds bind the semiconductor through a carboxylic acid group (either simple or as a part of a cyanoacrylic function).[43] However, the aqueous conditions employed for H$_2$ production and the different pH levels associated with different hole scavengers (*see below*) motivate the look for alternatives. A systematic study was conducted by Reisner and co-workers, who compared the performances of perylene monoimide (PMI) dyes endowed with different anchoring groups (Figure 3a, Table 1) in acidic, neutral and basic conditions, using two hole scavengers (triethanolamine and ascorbic acid).[44] Their main finding was that while a dye bearing the carboxylic group was very active and sufficiently stable under acidic conditions, moving towards higher pH the use of a phosphonic acid anchor became clearly preferable; interestingly, a dye with a hydroxyquinoline anchoring group (**PMI-HQui**) proved also very efficient under acidic conditions, but underwent fast deactivation as a result of detachment from TiO$_2$ surface.

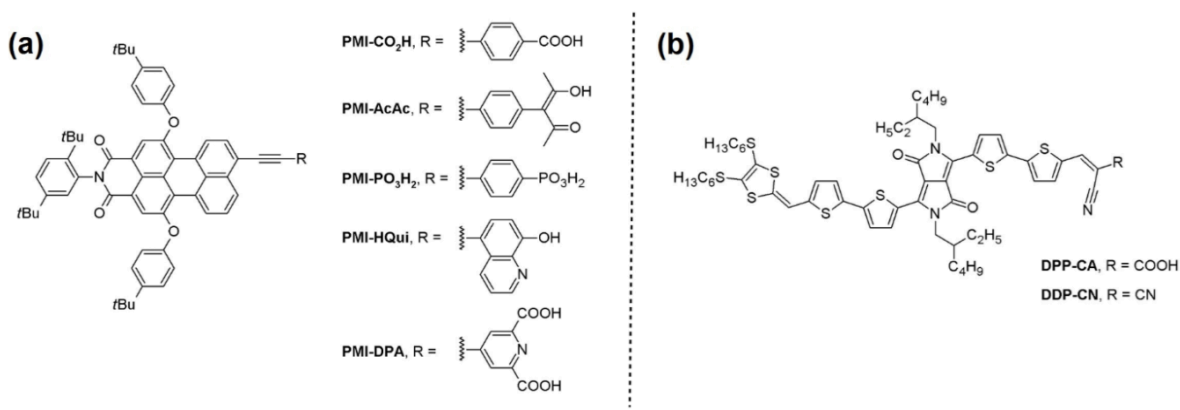

*Figure 3. Structures of organic dyes with different anchoring groups.*

**Table 1.** H$_2$ generation efficiency of PMI dyes with different anchoring groups.[44]

| Dye | Dye loading [µmol/g] | Conditions (SED, pH)$^a$ | H$_2$ produced [µmol] (24 h) | TON (24 h) | Retained activity after 24 h$^b$ |
|---|---|---|---|---|---|
| **PMI-CO$_2$H** | 13.3 | AA, pH 4.5 | 53.7 ± 6.2 | **6461 ± 749** | 78% |
| | | TEOA, pH 7.0 | 3.9 ± 0.5 | **471 ± 63** | 33% |
| | | TEOA, pH 8.5 | 4.1 ± 1.4 | 490 ± 170 | 35% |
| **PMI-AcAc** | 16.2 | AA, pH 4.5 | 21.7 ± 2.2 | 2146 ± 203 | 80% |
| | | TEOA, pH 7.0 | 1.3 ± 0.1 | 133 ± 13 | 51% |
| | | TEOA, pH 8.5 | 3.0 ± 0.7 | 294 ± 67 | 52% |
| **PMI-PO$_3$H$_2$** | 19.3 | AA, pH 4.5 | 42.5 ± 6.3 | 3546 ± 523 | 70% |
| | | TEOA, pH 7.0 | 3.6 ± 0.4 | 303 ± 30 | 46% |
| | | TEOA, pH 8.5 | 8.5 ± 1.3 | **708 ± 107** | 48% |
| **PMI-HQui** | 17.3 | AA, pH 4.5 | 53.3 ± 5.9 | 4928 ± 549 | 44% |
| | | TEOA, pH 7.0 | 2.5 ± 0.5 | 232 ± 26 | 38% |
| | | TEOA, pH 8.5 | 2.8 ± 0.4 | 262 ± 36 | 41% |

| | | AA, pH 4.5 | 41.4 ± 2.9 | 3943 ± 394 | 54% |
| --- | --- | --- | --- | --- | --- |
| **PMI-DPA** | 16.8 | TEOA, pH 7.0 | 3.8 ± 0.2 | 366 ± 37 | 55% |
| | | TEOA, pH 8.5 | 4.7 ± 0.7 | 444 ± 62 | 56% |

[a] 1.25 mg Dye/TiO$_2$/Pt in 3 mL 0.1 M SED solution, UV-filltered simulated solar irradiation (AM 1.5 G, 100 mW cm$^{-2}$, λ > 420 nm, 25 °C). [b] Calculated by comparing TOF values after 1 h and after 24 h.

A peculiar result was reported by Singh *et al.*, who compared the efficiency of two photocatalytic systems obtained with analogous dyes bearing a cyanoacrylic or a malononitrile anchoring group (Figure 3b). Surprisingly, it was the latter (**DPP-CN**) that produced the better result in terms of H$_2$ production in typical conditions (dye loading 25 μmol/g, TEOA 10% vol. in H$_2$O, pH 7, 2.0 Sun irradiation, λ > 400 nm), with a TON of 9664 in 10 h (corresponding to 1208 μmol of evolved H$_2$) compared to 6720 recorded for **DPP-CA** (840 μmol of evolved H$_2$).[45] Although the authors did not provide details on the anchoring mode of malononitrile to TiO$_2$, it is supposedly similar to that of dicyanomethylene compounds reported as sensitizers for DSSC.[46,47] Nevertheless, the latter were used as Type-II sensitizers and have a much simpler molecular structure, and therefore the working mechanism is hardly comparable in the two cases: given the excellent results reported, it seems that a deeper investigation of dyes with malononitrile or related anchoring groups could be useful to shed light on their behavior and further improve their performances.

Another key point is that the sensitizer, after photoexcitation and charge injection, should be readily regenerated by the reductant present in solution (either water in WS processes or a SED). Once again, this process has been thoroughly characterized in DSSC, and the main properties that a dye must possess to undergo efficient regeneration by a certain redox mediator are known in sufficient detail (driving force of the reaction *i.e.* dye HOMO position,[48] presence of certain functional groups on the donor section[49]). In the case of DSP systems, the situation is much less clear: an obvious requirement is for the sensitizer to have a more positive ground-state oxidation potential ($E_{S+/S*}$) than the standard redox potential of the reducing agent. However, this condition is not met in every case: for example, TEOA redox potential is reported to be +0.82-1.07 V *vs.* Normal Hydrogen Electrode (NHE),[17] but an efficient H$_2$ production was found also when employing it as a SED in combination with dyes having oxidation potentials in the +0.64-0.74 range.[45,50] Conversely, despite an apparently appropriate driving force for regeneration, many organic dyes with triphenylamine donors were found inactive when used together with ethanol as a hole scavenger.[51] Such apparent contradiction is probably due to two main reasons: first, dyes $E_{S+/S*}$ values are usually measured on diluted organic solutions in CH$_2$Cl$_2$ or CH$_3$CN, a very different environment compared to that in which they are actually used (adsorbed on TiO$_2$ in aqueous environment); second, different hole scavengers may work according to different reaction mechanisms and their redox potentials usually vary with pH,[17] which affects dye regeneration rates and thus photocatalytic turnover frequencies. In addition, the relative dye hydrophobicity/hydrophilicity could also influence its regeneration process (*see below*). To better evaluate the ability of organic dyes to work in DSP systems, it would be therefore advisable to investigate their electrochemical properties in more detail and in conditions more relevant to their actual application.

Furthermore, the reference electrode against which potentials are measured and the formalism used to convert such values to orbital energies (*vs.* vacuum) should be clearly indicated, as incomplete information often hinders comparison of data reported in different studies.

As a final point of this paragraph, spatial organization of dye molecules on the semiconductor surface should also be precisely controlled, to maximize the photocatalyst light absorption ability and reduce losses due to energy dissipation. Such a parameter has also often been associated with the relative hydrophobicity/hydrophilicity of the dyes, controlling their interactions with the solvent and the semiconductor surface (*see below*). Ahn, Son and co-workers found that by decorating phenothiazine dyes with alkyl chains of different length (Figure 4, **P1-P5**) a macroscopic effect on $H_2$ production efficiency could be observed, with the best result provided by dye **P5** featuring the largest substituent (TON after 5 h increasing from 380 for **P1** to 1026 for **P5**); the authors claimed that "*alkyl groups on nitrogen can induce a favorable orientation of dyes on TiO₂, which may result in the efficient electron injection from excited dyes to TiO₂*".[52]

Such a concept was further developed by Abbotto, Fornasiero and co-workers, who modified the same class of dyes and placed different hydrophobic and hydrophilic chains on the nitrogen atom (Figure 4).[53] They found that dye **PTZ-ALK**, featuring an *n*-octyl chain, showed a much higher $H_2$ production efficiency compared to its hydrophilic counterparts (**PTZ-TEG** and **PTZ-GLU**) at low dye loading, but such a difference was largely reduced when the loading was increased up to 30 μmol g$^{-1}$ (Table 2). According to the authors, at high loadings the organization of **PTZ-GLU** is "*similar to that of **PTZ-ALK**, with the PTZ units interacting with the Pt/TiO₂ surface and the bulky lateral chains avoiding intermolecular quenching*"; when the loading is decreased, though, "*the glucose unit could interact directly with the TiO₂ surface through the remaining OH groups and it might change the orientation of the PTZ scaffold affecting the electron transfer to TiO₂*".[53]

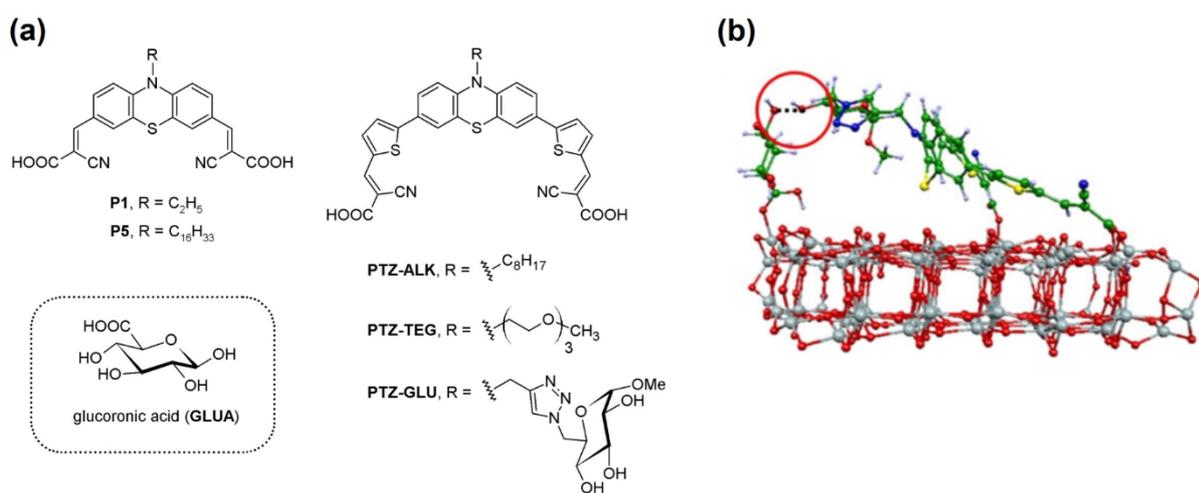

*Figure 4. (a) Structures of phenothiazine-based photosensitizers and **GLUA** additive; (b) DFT computational analysis showing the H-bond interaction between dye **PTZ-GLU** and **GLUA** (in the red circle). Reprinted with permission from reference* [54] *(© 2018, American Chemical Society).*

The importance of having a correct disposition of sensitizers molecules on the semiconductor surface was later confirmed when the same group studied the effect of combining **PTZ-GLU** with different co-adsorbents, including in particular glucoronic acid (**GLUA**, Figure 4).[54] It was found that using the sensitizer and **GLUA** in a 1:1 ratio clearly increased the TON of the photocatalyst; remarkably, this was not the case when a different and more hydrophobic co-adsorbent (chenodeoxycholic acid, **CDCA**) was used (Table 2). By employing DFT computational analysis, the authors found that a "*directional and selective*" interaction was established between **PTZ-GLU** and **GLUA**, which helped stabilizing the dye-semiconductor assembly and was effective in hindering dye-dye intramolecular interactions, minimizing unproductive energy transfer phenomena. This was confirmed by the fact that, when **PTZ-ALK** was combined with either **CDCA** or **GLUA** in the same ratio, no improvement was observed, since no selective interaction could be established between the coadsorbents and the bare alkyl chain.

**Table 2.** Photocatalytic data for **PTZ** dyes in combination with different coadsorbents.[53,54][a]

| Dye | Dye loading [µmol/g] | Coadsorbent | $H_2$ produced [mmol/g] (20 h) | TON (20 h) | $LFE_{20}$ [%][b] | AQY [%][c] |
|---|---|---|---|---|---|---|
| **PTZ-GLU** | 1 | - | - | 678 | 0.008 | - |
|  | 30 | - | 0.88 | 59 | 0.024 | 0.071 |
|  | 30 | **GLUA** (1:1) | 1.37 | 91 | 0.037 | 0.139 |
|  | 30 | **CDCA** (1:1) | 0.73 | 48 | 0.020 | 0.077 |
| **PTZ-ALK** | 1 | - | - | 1232 | 0.017 | - |
|  | 30 | - | 0.96 | 64 | 0.026 | 0.081 |
|  | 30 | **GLUA** (1:1) | 0.66 | 44 | 0.018 | 0.062 |
|  | 30 | **CDCA** (1:1) | 0.84 | 56 | 0.023 | 0.073 |
| **PTZ-TEG** | 1 | - | - | 396 | 0.005 | - |
|  | 30 | - | 0.421 | 29 | 0.013 | - |

[a] Conditions: TEOA 10% v/v solution in $H_2O$, pH 7.0, 20 h irradiation, visible light (λ > 420 nm). [b] $LFE_{20}$: light-to-fuel efficiency after 20 h; for details on its calculation, see ref. [32]; [c] obtained with light irradiation at 450 nm.

In summary, although dye design principles for DSSC and DSP may be similar, sensitizers for the latter application should be developed in response to specific requirements to allow performance improvements. While maintaining a wide and intense light absorption in the visible spectrum, charge injection rates into the SC should be improved, for example by investigating new anchoring groups exploiting unusual charge transfer mechanisms. Dye regeneration rates should also be enhanced by exact tuning of the sensitizers HOMO levels towards use with a specific electron donor; this operation should be assisted by measuring the dyes electrochemical properties under more realistic conditions, and by performing time-resolved spectroscopic analysis of dye regeneration by different species, as already done in DSSC.[55] Finally, dye organization on the SC surface should be optimized by exploiting the formation of ordered supramolecular structures, either using the dyes alone or by interaction with co-adsorbent species, not limited to those traditionally employed in DSSCs.

## 5. Hydrophobicity *vs.* hydrophilicity of the photocatalyst surface

Variation of the dyes relative hydrophobicity and hydrophilicity can have a significant impact on the photocatalyst performances. However, the simple question if it is better, in terms of $H_2$ production efficiency, to use a more hydrophobic or hydrophilic sensitizer has not yet been definitively answered. Clearly, dye optimization should not only aim at improving the individual properties of the molecules (structural, spectroscopic, electrochemical), but should also take into account the specific conditions in which the photocatalytic reaction is conducted, including solvent, pH, presence and nature of a hole scavenger, type of illumination and so on; in this regard, it is then possible that the optimal sensitizer in one case will be outperformed by a different compound in another, if the reaction conditions are not the same.

As mentioned above, some early studies examined the effect of placing alkyl chains of different lengths on the donor section of the sensitizers, usually finding that photocatalysts based on dyes with long (up to $C_{16}$) substituents provided the best results.[52,56,57] In a further refinement, the issue of where it is best to place such hydrophobic groups has also been recently investigated. Although comparing reports on different classes of dyes is not always straightforward, it has emerged that putting alkyl chains on the middle part of the organic dye structures can also be advantageous,[58,59] and even lead to enhanced results, as shown in Figure 5a, where the TON values for dyes **MB25** and **AD418** are compared to that obtained for dye **DF15** (TEOA as SED, pH 7).[51] Such an effect was attributed to a more efficient shielding of $TiO_2$ coupled with a higher dye regeneration rate, due to the lack of steric bulk on the donor group. Indeed, further increase of the dyes hydrophobicity by installation of alkyl chains *both* on their donor and intermediate sections can even be detrimental, as exemplified by the data collected for Pt@$TiO_2$/**OB1-3** photocatalysts (Figure 5b): after an initial improvement going from **OB1** to **OB2**, performances with the **OB3**-based system were almost back to the initial level, probably as a consequence of an excessive steric bulk and non-optimal interaction with the hole scavenger.[60]

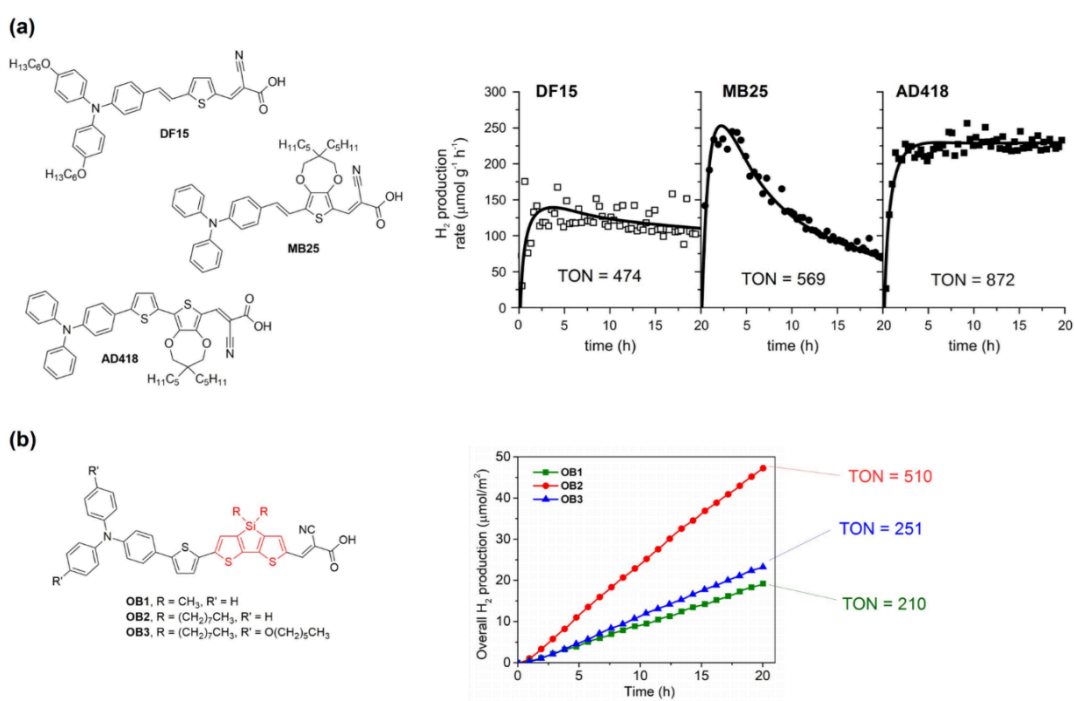

*Figure 5. Structures of (a) dyes **DF15**, **MB25**, **AD418** and (b) **OB1-3**, and the $H_2$ production curves of the corresponding Pt/TiO$_2$ dye-sensitized photocatalysts in the presence of TEOA as hole scavenger. Adapted with permission from references* [51] *(© 2018, Wiley-VCH) and* [60] *(© 2019, American Chemical Society).*

Moving to the opposite direction in terms of dye polarity, Kang and co-workers investigated the impact of changing the hydrophilic and steric properties of a series of organic dyes in sensitized H$_2$ generation using Pt/TiO$_2$ photocatalysts (Figure 6).[61,62] When using EDTA as sacrificial donor at acidic pH, it was found that hydrophilic methoxymethyl substituents at the 4,4′-positions of the diphenylamino end group enhanced the photocatalytic activity compared to both the parent compound (without substituents) and a hydrophobic counterpart. Differently from what seen above for hydrophobic chains, introduction of hydrophilic substituents also in the middle conjugated section of the molecules did not bring any further improvement. By applying both transient spectroscopy techniques and DFT calculations, the authors concluded that the different performances of the photocatalysts were due to a different organization of solvent molecules around the hydrophilic or hydrophobic substituents, coupled with steric effects that determined the amount of dye adsorbed on the semiconductor surface, which collectively influenced the kinetics of charge transfer processes across the SED/dye/semiconductor interfaces. For the best sensitizer, **MOD**, an AQY value of 0.27 ± 0.03% was measured under monochromatic light irradiation at 436 nm.

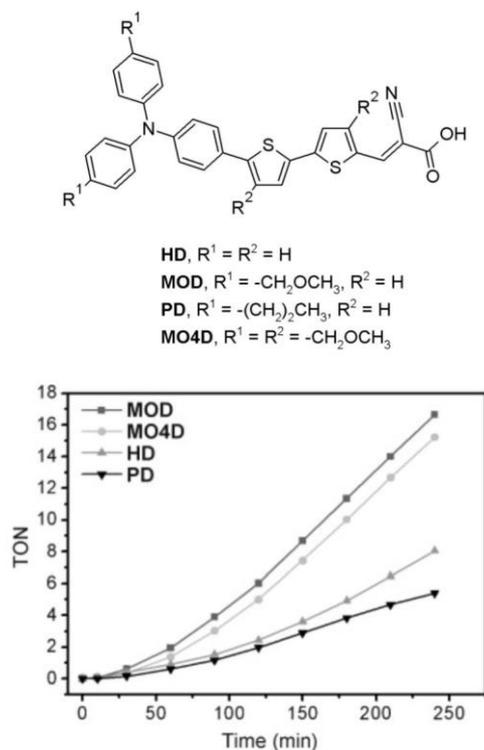

*Figure 6. (top) Structures of the hydrophilic sensitizers designed by Kang and co-workers, and of the corresponding hydrophobic dye. (bottom) TON of the H$_2$ production reaction of the corresponding Pt/TiO$_2$ DSP in the presence of EDTA as a hole scavenger. Reproduced with permission from ref.* [62] *(© 2012, Wiley-VCH).*

Despite their obvious interest, the above results turned out to be quite specific, as shown by the previously mentioned work by Abbotto, Fornasiero and co-workers on phenothiazine dyes,[53] in which they reported that dyes with hydrophilic substituents on the donor section were actually less efficient than that featuring a simple alkyl chain; it should be noted that their experiments were conducted in water at neutral pH and using TEOA as SED, thus in very different conditions compared to those performed by Kang *et al.*

The importance of correctly matching the sensitizers structure with the actual reaction conditions was further demonstrated in a recent study, in which a series of ten organic dyes based on the benzothiadiazole (BTD) core and featuring a different number of hydrophobic and hydrophilic chains were used as sensitizers for Pt/TiO$_2$ in H$_2$ production experiments with three different hole scavengers (TEOA, ascorbic acid, EtOH), at different pH levels (Figure 7).[63] The best performances with TEOA at pH 7 were obtained with highly hydrophobic dyes **BB2a** and **BB2d** (TON up to 295), whereas introduction of hydrophilic substituents on the donor section did not bring any improvement. Remarkably, when employing ascorbic acid as SED at pH 4, the situation was significantly changed, with the highest H$_2$ amount produced by the photocatalysts based on hydrophilic dyes **BB2e** and **BB3e** (TON up to 2266). Although the exact reason for the reversal in relative performances is not known, the improved interaction of the hydrophilic dyes with the polar SED molecules, coupled with a better matching of their ground-state oxidation potentials (compared to TEOA) and the easier proton reduction at lower pH clearly contributed to the observed outcome.

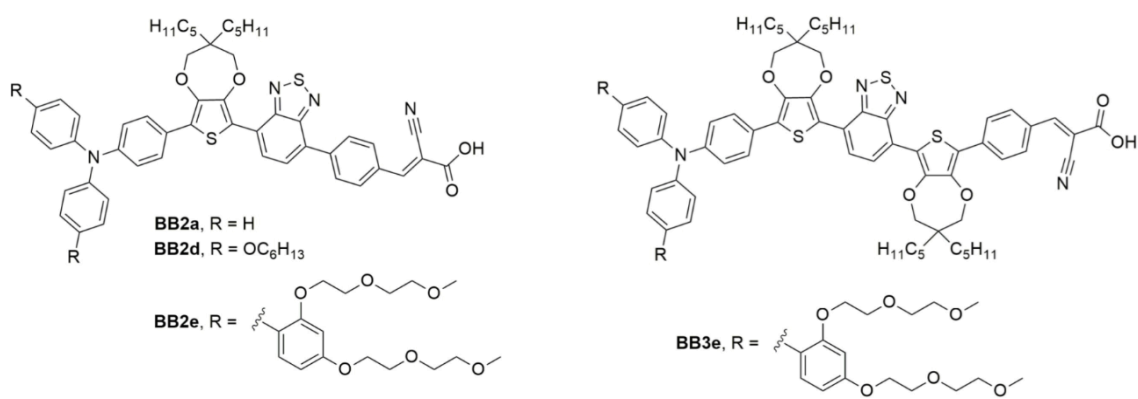

|  | Sensitizer | | | |
|---|---|---|---|---|
| TON Values | BB2a | BB2d | BB2e | BB3e |
| SED TEOA (pH 7) | 295 | 238 | 126 | 147 |
| SED AA (pH 4) | 1231 | 780 | 2266 | 1917 |

*Figure 7. (top) Structures of the hydrophobic and hydrophilic sensitizers of the **BB2** and **BB3** series; (bottom) TON values obtained in combination with different SEDs at different pH (dye@P25/Pt photocatalysts, dye loading 10 μmol/g, λ> 420 nm, irradiation time 15 h).*

In general, the results of all the above-mentioned studies suggest that no such thing as an "ideal dye" for DSP generation of H$_2$ exists, and that optimization of sensitizers properties must always be assessed in relation to the specific reaction conditions applied. In particular, choice of the sacrificial donor and of the pH level at which the reactions are conducted appear especially decisive in determining the H$_2$ production efficiency. In this context, it will be imperative in future years to improve dye design by introducing on the donor section functional groups able to interact efficiently with the selected SED molecule, which, together with appropriate tuning of the energy levels (see previous section), should help increase regeneration rate constants. To this end, we think that investigation of dyes with combined hydrophobic/hydrophilic sections should be continued, trying to favor structures able to provide a significant hydrophobic barrier against recombination near the SC surface, while at the same time bearing hydrophilic groups of carefully tuned steric bulk near the region where interaction with SED is thought to happen.

## 6. Effect of dye loading on photocatalytic performances

The effect of dye loading on photocatalyst performances is usually evaluated in two different ways, either by saturation of the semiconductor surface with dye molecules or by adsorption of a precise amount of sensitizers. In the first approach, the semiconductor nanoparticles are suspended in a solution containing a large amount of dye, so that adsorption is maximized: as a consequence, different sensitizers will be adsorbed in different amounts, depending mostly on their size and on their geometrical properties. In this way, it is possible to evaluate the relative $H_2$ production abilities of the corresponding photocatalysts, but no precise information on the individual dye efficiency and on its optimal loading can be obtained.

For example, this was the case in the above-cited work by Liu *et al.*,[59] where the maximum possible amount of "starbust" dyes **DH1-4** was loaded on Pt/mc-TiO$_2$ (an especially-developed anatase cubic "microcage" TiO$_2$ material). The TONs registered after 20 hours for dyes **DH3-4** were higher than that obtained for dye **DH2**, but the latter had a much higher adsorption density on the semiconductor surface, and thus the corresponding photocatalyst produced a higher $H_2$ amount (Figure 8 and Table 3).

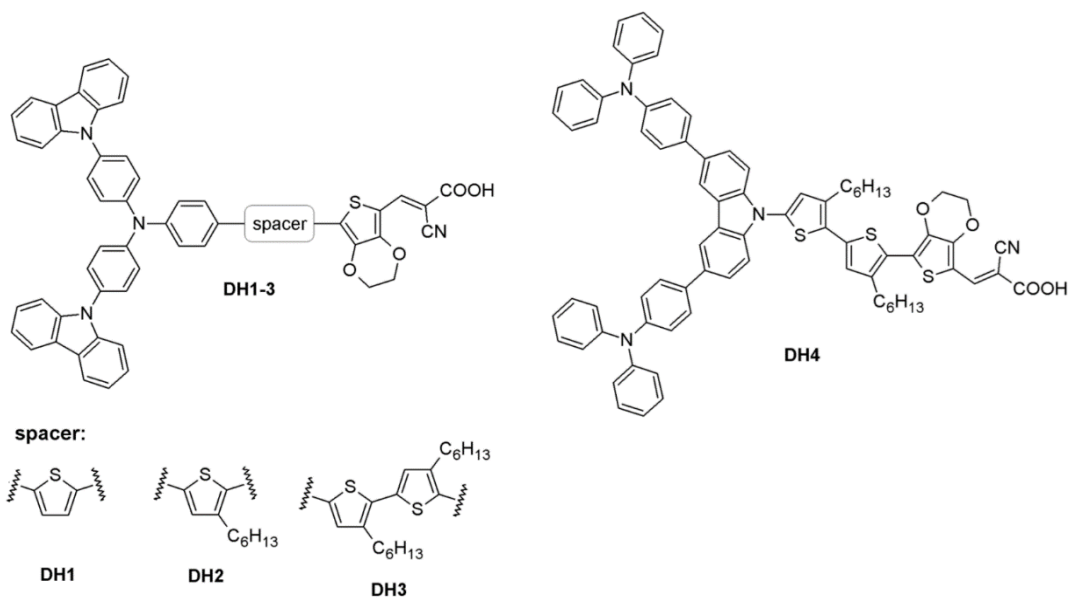

*Figure 8. Structures of dyes **DH1-4**.*[59]

**Table 3.** Photocatalytic data for **DH1-4**/Pt/mc-TiO$_2$ three-component systems.[59][a]

| Dye | Dye loading [µmol/g] | H$_2$ produced [mmol/g] (20 h)[b] | TON (20 h)[b] |
| --- | --- | --- | --- |
| **DH1** | 47.18 | 46.42 | 984 |
| **DH2** | 59.11 | 84.75 | 1434 |
| **DH3** | 47.80 | 75.68 | 1583 |
| **DH4** | 32.77 | 74.88 | 2285 |

[a] Conditions: 50 mg of dye/Pt/mc-TiO$_2$ in 100 mL of a 10% TEOA solution in H$_2$O. Irradiation with a 300W Xe lamp equipped with a cut-off filter (λ > 420 nm). [b] Experiments were run in triplicate, only the best result is shown.

In the second approach, a solution containing a precise quantity of sensitizer is used to stain the semiconductor, so that after sensitization a colorless supernatant solution is obtained. In this way, a precise amount of dye can be loaded on the photocatalyst, allowing to determine the effect of different dye loadings and to assess the relative TON values of the dyes at the same level of superficial concentration. A common finding in this kind of experiments is that the overall amount of generated H$_2$ initially increases with the increasing dye loading, but then reaches a maximum and starts decreasing again above a certain dye concentration. For example, such an effect was observed in several studies by Pal *et al.* examining organic sensitizers with different structures and anchoring groups,[45,50,57,64] and was attributed to the fact that initially the fraction of incident light absorbed by the dye increases with increasing dye loading, but then the photocatalytic activity starts to decline due to dye aggregation (accompanied by unproductive intermolecular energy transfer) and shielding effects reducing the penetration depth of incident light.

Clearly, since the maximization of dye loading on the semiconductor surface does not always lead to improved performances, this second approach appears preferable and a screening of the effect of dye concentration is recommended to obtain photocatalysts with optimized efficiency.

## 7. Impact of TiO$_2$ crystal structure: which phase is the best?

In the case of DSP, the semiconductor basically acts as an electron transporter from the sensitizer to the catalytically active site, so that, as mentioned above, one of the main requirements to be met by the sensitizer is that the injection of the excited electron into the semiconductor conduction band (CB) is allowed. Usually, such thermodynamic restriction is condensed into the need for "correct band alignment", namely the potential energy of the LUMO level of the dye must be more negative than that of the semiconductor CB. The widespread prominence of TiO$_2$ is often jeopardized by critical factors that compromise its performance, and decrease it to an unacceptable level. From the point of view of the semiconductor, some of the setbacks, particularly pronounced for first row transition metal oxides, include a fast charge recombination, poor charge mobility, surface effects, size of the band-gap and others.[65,66] One additional aspect is the relationship between the metal oxide crystal structure and the performance, which can be usefully exploited for better photocatalyst design. TiO$_2$, which is stable under ambient conditions in the three polymorphs rutile, anatase and brookite, is the best case study to fathom such a relationship. The three phases exhibit different photocatalysis-relevant properties such as charge recombination rate, band gap, density if states (DOS) and mode of charge carrier transport. Although such differences have been correlated to variations of photocatalytic efficiency,[67] establishing an univocal trend is a complex matter, because there is a strong dependence on the nature of the catalyst, being a single crystal, a thin film or a powder.[68] For example, while a recent study conducted on different anatase, brookite, and rutile single-crystal wafers with only one exposed surface showed that the anatase surfaces are generally more active than those of rutile and brookite for methanol photooxidation,[69] investigation of composite materials for alcohol photoreforming revealed that the hydrogen production relative to the surface area increased with brookite content, suggesting that brookite facets were more active for proton reduction under those conditions.[70]

Based on the catalyst nature, the presence, type and distribution of defects plays a very important role, whereby conduction band electrons can be trapped and stabilized to different extents, with the specific TiO$_2$ crystal structure being a powerful determinant.[71] Furthermore, additional factors come into play as well, such as complex charge transport kinetics within TiO$_2$[72] or varying particle size distribution,[73] making it difficult to shape a comprehensive and reliable paradigm related to predicted activity of each material.

Despite such a complexity, the built knowledge on TiO$_2$ crystal structure/photocatalytic dependence has resulted in very interesting new outputs, arising from the wise exploitation of advanced techniques. For instance, the once overlooked brookite, long considered an inactive phase, has recently gained attention due to its peculiar physico-chemical properties,[74,75] whose assessment was made possible by the emergence of new strategies for its synthesis in a pure form.[76] In the context of DSP, it was recently demonstrated that photocatalysts obtained by sensitization of nanocrystalline brookite/Pt with the above-mentioned sensitizer **OB2** (Figure 5) provided better performances in H$_2$ production experiments compared to their P25-based counterparts (Figure 9a), being also characterized by a remarkable stability (Figure 9b).[60] This result was attributed to a reduction in charge recombination rate due to the lower reactivity of conduction band electrons of brookite compared to anatase,[71] in agreement with previous studies conducted on DSSCs.[77]

In addition, the morphology of the TiO$_2$ is to be taken into careful consideration. Several groups have reported the use of anatase-based semiconductors with tailor-made morphology for use in DSP systems, such as cubic "microcage" materials[59] or hierarchical porous structures,[31,39,50] showing enhanced performances compared to the commercial TiO$_2$ sources, owing to improved electronic features or larger surface area. Cargnello *et al.* demonstrated how the geometrical anisotropy of brookite nanorods was instrumental for improving charge separation, with the possibility to tune the photocatalytic activity for H$_2$ evolution by controlling the nanorods length.[78]

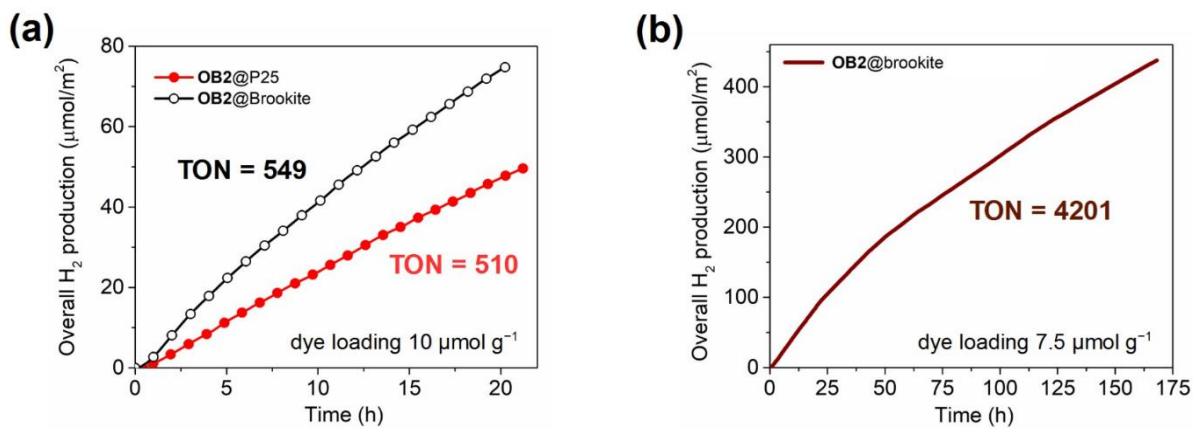

*Figure 9. (a)* H$_2$ *production per photocatalyst surface area of* **OB2**-*sensitized P25/Pt (red circles) and brookite/Pt (black hollow circles) over 20 h visible light irradiation (λ > 420 nm, dye loading, 10 μmol·g$^{-1}$); (b)* H$_2$ *production relative to surface area of* **OB2**@*brookite/Pt photocatalyst over 170 h of visible light irradiation. All experiments were performed with TEOA as SED. Reproduced with permission from ref.* [60] *(© 2019, American Chemical Society).*

In view of the interesting results already obtained, investigations on the combination of dyes with different TiO$_2$-based materials should be continued. In particular, studies should concentrate on the development of semiconductors with tailored morphology to speed up charge transport and transfer to the HEC, while minimizing charge recombination rates. In addition, studies should be conducted on the sensitization of polymorph mixtures other than the common P25, such as for example brookite/anatase mixtures, to take advantage of both the higher reactivity towards hydrogen reduction and the enhanced degree of charge separation thanks to the presence of phase boundaries.

## 8. Approaches to improve stability

Being as important as activity, the photocatalyst stability requires attention, and when optimizing a system for H$_2$ production all possible phenomena contributing to its deactivation should be investigated. In the specific case of DSP, the typical deactivation mechanisms observed in non-sensitized SC photocatalysts, such as surface passivation or photocorrosion processes,[79] can be accompanied by additional sensitizer-related degradation pathways, which can be related both to the strength of their bond with the SC and their intrinsic chemical and photochemical stability.

First, photocatalyst deactivation can occur due to partial or complete detachment of the dye from the semiconductor surface, which clearly depends on the kind of anchoring group placed on the sensitizer structure. We have already alluded to this aspect when discussing the work of Reisner *et al.* on PMI dyes endowed with different anchoring groups (*see above*),[44] although there we mostly focused on photocatalytic performances. In general, it has been reported that the carboxylate linkage may not be an optimal choice when employing dye-sensitized photocatalysts in aqueous environment, due to accelerated hydrolysis of the titanate ester linkage, especially at basic pH. For this reason, the use of more robust anchors, such as phosphonate derivatives, has become increasingly popular, although it is still more common for Ru-based organometallic dyes [80,81] than for metal-free organic structures.[82] In this regard, an interesting alternative could be represented by the use of a silane coupling reagent to covalently anchor the sensitizer to TiO$_2$: such approach was demonstrated in a seminal paper by Arakawa and co-workers, who reported that chemical fixation of Eosin Y through amide coupling with Pt/TiO$_2$ functionalized with γ-aminopropyl-triethoxysilane yielded a stable and efficient photocatalytst for H$_2$ production from TEOA (Figure 10).[83] To the best of our knowledge, such strategy has not been applied further in DSP systems, although dyes with silane and silatrane anchors were later used to sensitize metal oxide electrodes for photoelectrochemical cells,[84,85] and were shown to give DSSCs with high power conversion efficiencies.[86] Further studies could help establish how the siloxane anchoring group should be connected to the sensitizer structure to provide optimal charge transfer rates, a matter that has undergone in-depth scrutiny in the field of DSSC.[87]

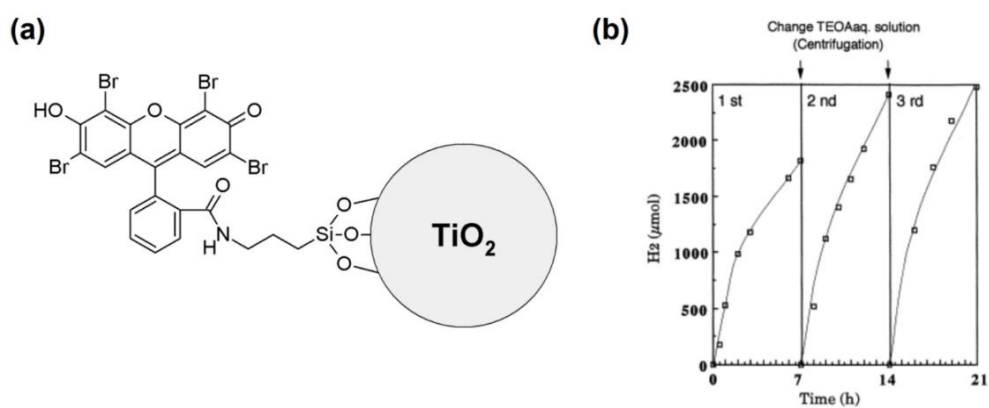

*Figure 10. (a) Eosin Y anchored to TiO$_2$ through an aminosiloxane linker; (b) stable H$_2$ production along consecutive photocatalytic experiments. Reproduced with permission from ref. [83] (© 2000, Elsevier B. V.).*

Furthermore, the anchoring stability of the dye can be improved by increasing the number of anchoring groups as studied in detail by Park and co-workers, who prepared three triphenylamine-based sensitizers, **D1-3** bearing one, two or three cyanoacrylic anchoring groups, respectively, and studied their possible binding modes on TiO$_2$ (Figure 11a).[88] Although *in situ* IR studies suggested that simultaneous binding of all three carboxylic acids was hardly possible, dyes **D2** was observed to give both *mono-* and *bis-*coordinated complexes on TiO$_2$, while **D3** was bound mostly in *bis-*coordinated fashion. Consequently, in photocatalytic experiments dyes **D2-3** gave better efficiency and stability compared to **D1** probably as an effect of their more robust anchoring on TiO$_2$ (Figure 11a). Similar observations were made in the already-cited work by Son *et al.*, who observed that bidentate phenothiazine dyes yielded consistently better performances compared to their analogues with only one anchoring unit (Figure 11b),[52] as well as in a study by Watanabe, Tani and co-workers, investigating porphyrin derivatives with mono- or multi-pyridyl anchoring groups.[89]

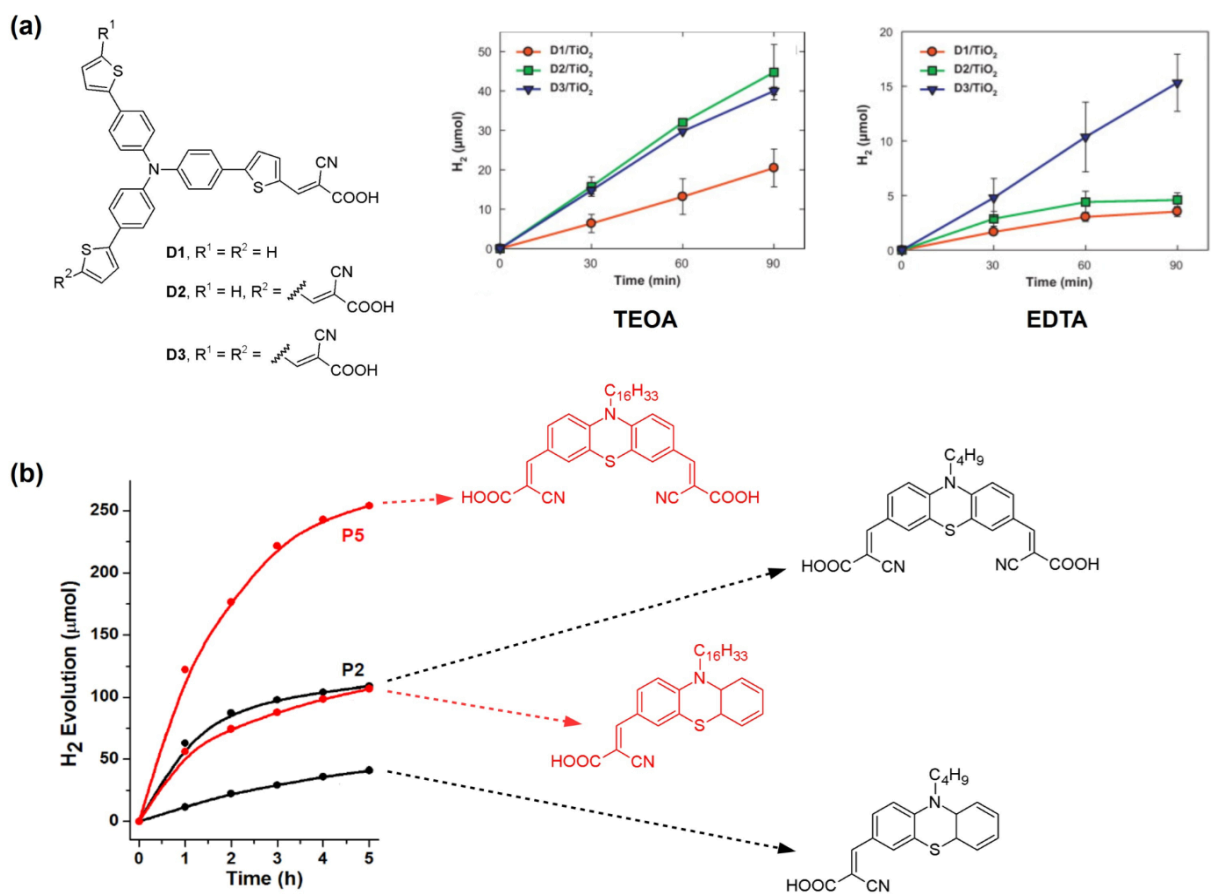

*Figure 11.* *(a) mono-, bis- and tris-cyanoacrylic dyes **D1-3**, and the performances of the corresponding photocatalysts in $H_2$ evolution experiments with TEOA and EDTA as sacrificial donors. Reproduced with permission from ref.* [88] *(© Elsevier B. V., 2012); (b) Photocatalytic data for mono- and bis-coordinating phenothiazine dyes. Reproduced with permission from ref.* [52] *(© 2012, Royal Society of Chemistry).*

Another key aspect to consider to enhance photocatalyst stability is preventing the intermolecular quenching that follows agglomeration of the dye molecules. A common approach to solve the issue, often an indispensable requirement, is to endow the dye molecule with an encumbered steric environment; indeed, it has been repeatedly demonstrated that placing alkyl chains of sufficient length in the intermediate section of the sensitizers can provide the necessary steric bulk to avoid dye aggregation.[51,59,60,64,90] A remarkable example was provided by Abbotto, Fornasiero and co-workers, in their work on $H_2$ production by DSP featuring phenothiazine dyes (Figure 12).[91]

They found that, at the beginning of the photocatalytic experiment dye **PTZ1** provided a better performance than all other analogues named **PTZ2-6**; however, after a prolonged period of time, the overall amount of gas produced by dye **PTZ5** was higher, as a result of a superior photocatalytic stability, as visible by its constant $H_2$ evolution rate. Although the reasons for this result are not completely clear, the presence of *n*-butyl chains in the middle part of **PTZ5** surely helped to reduce dye agglomeration and limit undesired energy transfer processes between dye molecules. As mentioned above, another strategy to optimize the dye geometry on the SC surface and hinder dye-dye interactions is to use co-adsorbents, especially by exploiting the formation of directional hydrogen bonds with the sensitizer molecules,[54] but a definite effect on DSP stability for such systems has not been reported.

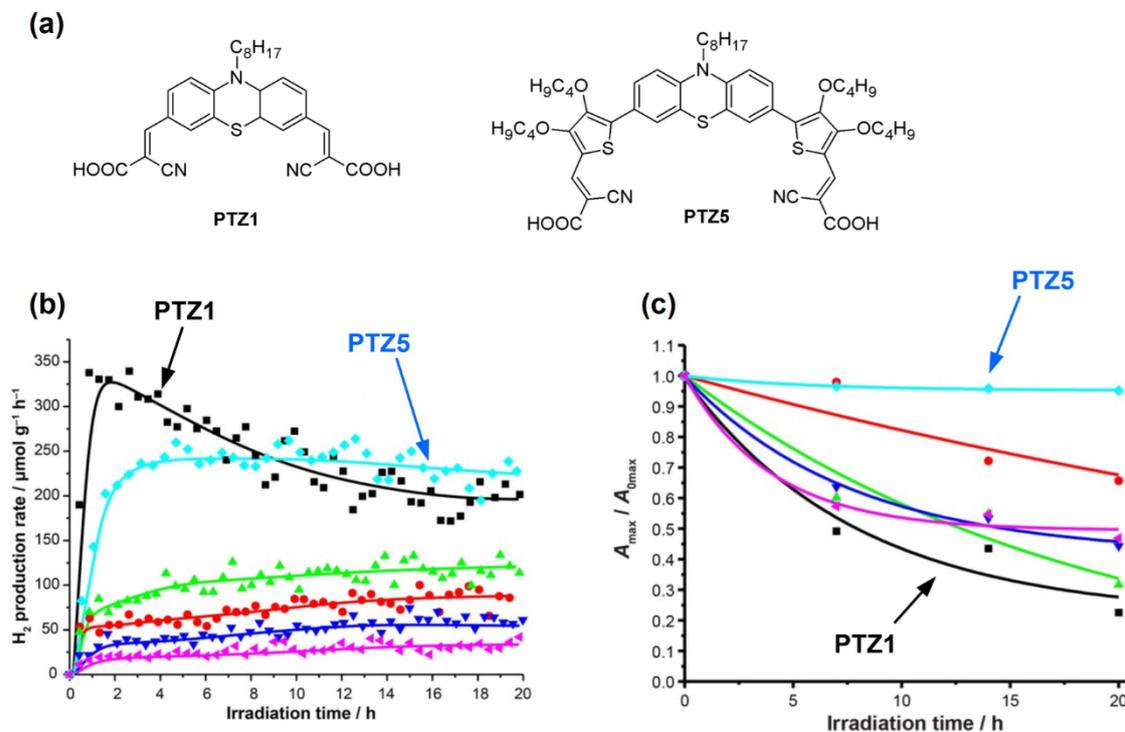

*Figure 12.* (a) Structures of sensitizers **PTZ1** and **PTZ5**; (b) $H_2$ production rates measured using the dye/Pt/TiO$_2$ materials suspended in TEOA 10% v/v solution at pH 7.0 under irradiation with visible light (λ>420 nm); (c) Degradation plots of the dye-sensitized Pt/TiO$_2$ photocatalysts under visible irradiation in the same conditions of $H_2$ production experiments. Reproduced with permission from ref. [91] (© 2015, Wiley-VCH).

The instability of DSP can also derive from the degradation of the dye over time by reaction with chemical quenchers present in solution or reactive species formed during photocatalysis, such as $H_2$ itself. Indeed, it was observed early,[83] and confirmed subsequently,[92] that emissive dyes such as Eosin Y can undergo irreversible hydrogenation by $H_2$ in the reaction conditions, giving species characterized by a lower degree of conjugation and as such less able to absorb visible light, thus hindering photocatalytic activity.

The different stability of the above-mentioned dyes **MB25** and **AD418** was interpreted in terms of their different resistance to degradation during photocatalysis. **MB25** presented an electron-donating propilenedioxythiophene (ProDOT) ring next to a double bond, which activated a decomposition pathway starting with dye protonation and nucleophilic attack by water; **AD418**, in which the double bond was substituted by a thiophene ring, could not undergo the same side reaction and therefore gave rise to a much more stable photocatalytic system, resulting in a far higher TON.[51] Finally, in a recent paper Lai et al. studied the degradation process of multicarbazole-based organic dyes to understand the issues related to stability of Pt/TiO$_2$ photocatalysts for $H_2$ evolution. Supported by combined UV–Vis, FT-IR, $^1$H and $^{13}$C NMR, and MS techniques, it was suggested that the decline of activity matched the progressive removal of the electron acceptor unit (consisting of cyanoacrylate moiety), via initial decarboxylation reaction followed by removal of the $CN^−$ group, a mechanism previously unreported for DSP systems.[93]

Thanks to the advances in dye design and in the investigation of photocatalyst deactivation pathways, several DSP systems, including some of those cited above, have been demonstrated to achieve prolonged stability in H$_2$ production experiments, with the best examples being still noticeably active after more than 100 h under continuous illumination (Table 4). Unfortunately, setup of such experiments is nontrivial, especially for those groups having access to only one photochemical reaction apparatus, and thus extended stability studies (at least > 48 h) are still lacking in some of the recently published works. Given the importance of the photocatalyst stability parameters, however, they appear indispensable for a complete and fair assessment of new DSP systems and should be always included whenever possible.

**Table 4.** Stability data of some selected DSP systems.

| Dye | Reaction Time (h) | Dye loading [μmol/g] | H$_2$ produced [mmol/g] | TON | SED (pH) | Ref. |
|---|---|---|---|---|---|---|
| **Alizarin** | 80 | 2.5 | 7.91[a] | 6326 | TEOA (9) | [92] |
| **Alizarin Red** | 92 | 2.5 | 7.93[a] | 6342 | TEOA (9) | [92] |
| **PTZ5** | 90 | 59.6 | -[b] | -[b] | TEOA (7) | [91] |
| **S1** | 48 | 6.25 | 63.75 | 10200 | AA (4) | [40] |
| **Dimer 2** | 83 | 28.3 | 84.91 | 2860 | AA (4) | [89] |
| **OB2** | 170 | 7.5 | 15.75 | 4201 | TEOA (7) | [60] |
| **Calix-3** | 50 | 37.3 | 630.97 | 16927 | TEOA (11.8) | [94] |
| **DH4** | 105 | 32.8 | 547.22 | 16699 | TEOA (n.d.[c]) | [59] |
| **BB3e** | 72 | 2.5 | 29.11 | 23285 | AA (4) | [63] |

[a] Calculated based on the TON and dye loading data presented in the original paper. [b] Exact values were not provided; Figure S9 in the supporting information of the original publication shows a constant H$_2$ production rate of approx. 250 μmol g$^{-1}$ h$^{-1}$ for the entire experiment. [c] The authors report that the solution pH was adjusted by addition of perchloric acid.

In view of the above discussion, improvement of DSP stability should be first pursued by making the dye/semiconductor assembly more robust. Accordingly, a more thorough exploration of structures with multiple binding sites to TiO$_2$ should be carried out. On the other hand, care shall also be placed in designing dyes not incorporating labile functional groups in their central section. In this regard, it will be preferable to prepare compounds with directly connected (hetero)aromatic rings, without the presence of multiple (double/triple) bonds, and without excessively electron-donating moieties, as they could be progressively oxidized during the H$_2$ evolution reaction.

## 9. Nature of the hydrogen evolution catalyst (HEC)

Usually, in dye-sensitized photocatalytic systems for H$_2$ production, proton reduction is carried out by metal nanoparticles adsorbed on the semiconductor surface, with platinum being by far the most common choice.[12]

Clearly, this is due to the excellent properties of platinum as a heterogeneous catalyst for $H_2$ evolution, guaranteeing high activity and stability, but also to the fact that most of the studies focus on the investigation of other components of the system (such as the dye or the semiconductor) and therefore need to use the same catalyst to obtain results comparable with those already reported in the literature.

Nevertheless, several studies have focused on finding more readily available and cheaper catalysts than platinum, in the perspective of an industrial scale-up of the system. Indeed, it has even been shown that $H_2$ production with dye-sensitized $TiO_2$ can proceed also in the absence of adsorbed metals,[95] but usually gas evolution rates were not sufficient for practical purposes. In addition, the possibility to use dissolved homogenous metal catalysts, not anchored to $TiO_2$, has also been explored: for example, Kruth *et al.* reported the employment of commercially available $PdCl_2(CH_3CN)_2$ and $Pd(PPh_3)_2Cl_2$ as catalysts in combination with polymer-capped titania nanoparticles sensitized with ruthenium complex **N3**.[96] Although a moderate and stable $H_2$ evolution was obtained, the authors mention that the results were inferior to those previously reported for other composite $TiO_2$ photocatalysts.

More commonly, transition metal or metal salt nanoparticles adsorbed on the semiconductor surface have been reported as catalysts for DSP systems. Although this has been done more often for purely inorganic photocatalytic assemblies,[97–99] several examples exist also in the dye-sensitized field. Already in 2007, Lu and co-workers described the use of an Eosin Y-sensitized $CuO/TiO_2$ nanocomposite, in which cuprous oxide played the double role of semiconductor and catalyst for water reduction, being able to collect electrons directly by injection from the sensitizer or through electron transfer from titania; its employment allowed to obtain a much higher $H_2$ production rate compared to that observed in its absence.[100] More recently, the use of elemental Co was also reported in a similar system, in which Rhodamine B was used as the sensitizer; remarkably, the authors reported the possibility to achieve a full water splitting process, avoiding the use of any sacrificial donor, thanks to the synergistic effect of the sensitizer and the neighbouring cobalt atoms. The fact that the reaction proceeded through the desired mechanism was supported by the production of nearly stoichiometric amounts of the two gases ($H_2:O_2$ ratio was approximately 2.3).[101] Du and co-workers examined several first-row transition metal-based oxide/hydroxide materials, such as cobalt oxide ($CoO_x$), cobalt hydroxide ($Co(OH)_2$), nickel oxide ($NiO_x$), nickel hydroxide ($Ni(OH)_2$), ferric hydroxide ($Fe(OH)_3$) and copper hydroxide ($Cu(OH)_2$), as catalysts in a three-component photocatalytic system with $TiO_2$ as the semiconductor, Eosin Y as the sensitizer and TEOA as SED (Figure 13). They found that $Ni(OH)_2$ exhibited the best performance, which was about 90 times higher than that of pure $TiO_2$ under the same conditions and was kept stable for several hours through repeated illumination/evacuation cycles.[102]

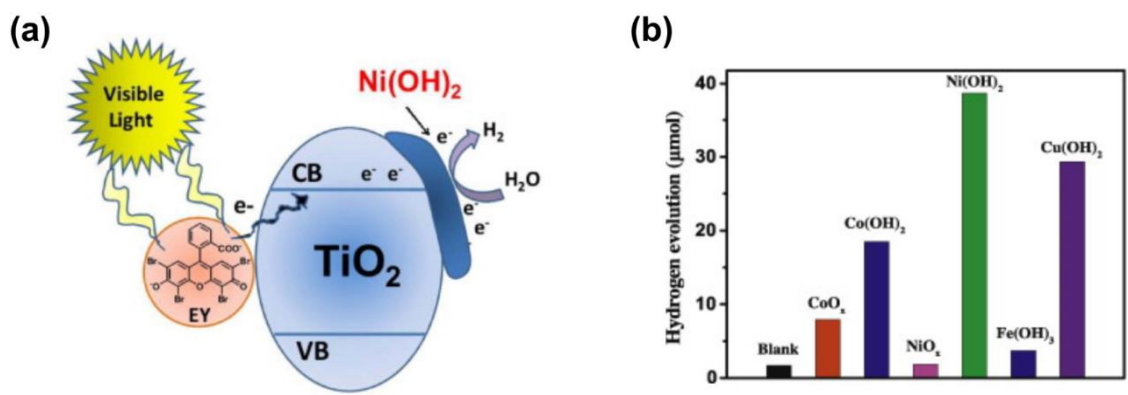

*Figure 13. (a) Mechanism of Eosin Y-sensitized H$_2$ production with Ni(OH)$_2$/TiO$_2$ nanoparticles; (b) comparison between different transition metal oxides/hydroxides used as HEC. Reproduced with permission from ref. [102] (© 2014, Elsevier B. V.).*

A series of studies was published by Patir and co-workers on the use of Cu$_2$WS$_4$ nanocubes as HEC in combination with TiO$_2$ sensitized by a variety of different metal-free organic sensitizers.[41,103–105] The new catalyst was synthesized by a hot injection method that produced nanocubic structures with 100-500 nm-long edges and characterized by a single crystalline phase. Photocatalytic studies revealed that use of Cu$_2$WS$_4$ caused an increase in the H$_2$ production rate compared to the catalyst-free dye-sensitized semiconductor, but its performances were lower than those of Pt nanospheres. The system was also sufficiently stable under irradiation, although XPS measurements conducted both before and after the experiments indicated a partial hydrogenation of the Cu$_2$WS$_4$ structure during the photocatalytic reaction.

Finally, it should be mentioned that several examples of supported molecular catalysts have also been reported in DSP systems for H$_2$ production. Many of these studies have been conducted by Reisner's group, who focused especially on the development of Co- and Ni-based complexes (Figure 14), used in combination with both Ru-containing sensitizers and metal-free organic dyes. Earlier work concerned the employment of cobaloxime complexes such as **CoP$^1$**,[106,107] whose attachment to the semiconductor was allowed only by an axial pyridine ligand endowed with a phosphonate anchoring group. Later, the catalyst design was improved by preparing complex **CoP$^3$**, featuring a single ligand incorporating both the diamine-dioxime equatorial unit and the axial pyridine:[108] accordingly, the authors reported that "*CoP$^3$ displays significant advantages over previously reported immobilized Co catalysts as it shows a higher catalytic proton reduction activity and provides a strong and more stable anchoring to metal oxides surfaces*". At the same time, DuBois-type [Ni(P$_2^{R'}$N$_2^{R''}$)$_2$]$^{2+}$ complexes were also investigated in combination with Ru *tris*(bipyridine) dyes, and it was shown that they could work in water reduction reactions both in homogenous phase or adsorbed on a semiconductor, albeit with different electron transfer mechanisms.[109] In general, the performances provided by these molecular catalysts were good, but inferior to those obtained with Pt nanoparticles.[44,82]

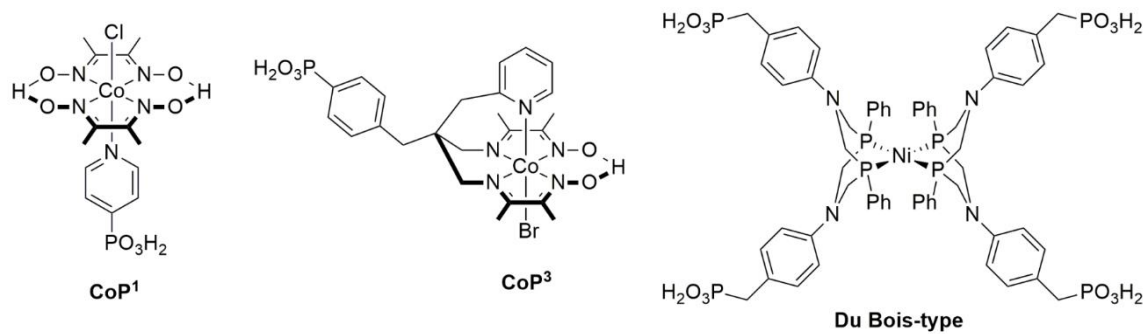

*Figure 14. Structures of molecular complexes used as HEC in dye-sensitized systems.*

Despite that, we consider studies on alternative HECs of high relevance in the perspective of a potential future large-scale deployment of DSP technology, and recommend that they will be expanded in the future also by other research groups. Finding reliable catalytic species based on cheaper and more available metals than Pt could significantly reduce the projected cost of DSP systems, while at the same time eliminate (or largely reduce) the risk of shortages of critical materials in the long term.

## 10. Sustainability of the sacrificial donor

As explained in the introduction, during photocatalysis on DSP (not differently from purely SC-based photocatalysts), charge separation following light irradiation generates two reactive sites, where the newly formed holes and electrons can promote oxidation and reduction reactions, respectively. The development of new photocatalysts is heavily based on the detailed understanding of its intrinsic activity. Hence, it is convenient to simplify the investigation of the photocatalysts features by focusing on one half-reaction only, relying on the use of sacrificial electron donors or acceptors (SED or SEA respectively) that readily react with the photogenerated charge carriers, thus not placing kinetic restrictions that would affect the half-reaction of interest.[110] For example, such a practice is common in the photocatalytic water splitting field, where most studies focus on either the $H_2$ evolution or on the water oxidation, using sacrificial agents for the other half-reaction. Focusing on the $H_2$ evolution process, obviously the choice of the SED is subject to stringent thermodynamic requirements for the reaction to proceed efficiently. Among them, the most important is that the HOMO of the dye must suitably match the redox potential of the $SED_{red}/SED_{ox}$ couple to ensure the rapid regeneration of the oxidized dye (*i.e.* the HOMO of dye must be more positive than the oxidation potential of the SED). Typical SED include triethylamine (TEA), triethanolamine (TEOA), ethylenediaminetetraacetic acid (EDTA), ascorbic acid (AA), $S_2^-$, $I^-$ and others.[17]

However, while alleviating the complexity demands of photocatalyst developments from fundamental perspective, this approach does not match sustainability concepts. To address this aspect of photocatalysis, an increasing load of research has moved towards more useful SED, whose oxidation can be associated with other processes relevant for sustainability.[111] For example, alcohols can act as efficient donors for many inorganic semiconductor photocatalysts including $TiO_2$,[112] with potential oxidation to industrially relevant compounds, as recently demonstrated for ethanol and glycerol.[113] Despite that, the use of alcohols as hole scavengers in DSP systems is still in its infancy, and has been reported only in a handful of studies, investigating the employment of MeOH,[114] EtOH[51,63] or glycerol.[50] Unfortunately, the performances obtained with such sacrificial donors are generally lower than those registered with TEOA or AA, and the structural and electronic requirements of the dyes to work efficiently with them have not yet been fully clarified. Probably, one key issue is that small alcohol molecules can adsorb on the SC surface, reducing the $H^+$ reduction rate and enhancing charge recombination;[115] therefore, they should be used in combination with dyes able to efficiently shield the photocatalyst surface but at the same time small and hydrophilic enough to allow good interaction with the SED. Achieving such a structural design is nontrivial and therefore studies on sensitizer optimization are still in progress. In addition, alcohol oxidation could be promoted by combining the dyes with appropriate molecular catalysts,[116] also in an integrated dyad design, as already demonstrated in photoelectrochemical cells.[117]

Another promising research direction would be to explore the photoreforming (PR) of biomass-derived materials, such as lignin and lignocellulose, as hole scavengers, opening the way to the production of clean fuels from abundant and cheap raw materials, or even waste.[118] Despite their favourable thermodynamics, however, such raw materials are often characterized by limited solubility, brown-dark colour and slow oxidation kinetics, making it necessary to apply pre-digestion procedures and use appropriate catalysts for their efficient photoconversion. Due to such issues, no DSP system for lignocellulose PR has been reported as of yet, but given its great potential such approach should definitely be pursued in the future.

## 11. Perspectives

In this article, we have highlighted some key aspects of the visible light-driven $H_2$ production mediated by heterogeneous dye-sensitized photocatalysts. Compared to the other two main technologies currently employed for the production of solar $H_2$, namely tandem photovoltaic-electrolysis systems and photoelectrochemical cells, the photocatalytic approach is still characterized by an inferior solar-to-hydrogen (STH) efficiency, but at the same time is comparatively simpler, less expensive and easier to scale-up.[4] However, to be able to replace, at least partially, $H_2$ generation methods based on more mature technologies (such as hydrocarbon reforming or water electrolysis), photocatalytic processes still need to overcome some serious obstacles. In general, the major factors currently restricting large-scale application of DSP systems towards a possible industrialization can be traced back to their insufficient efficiency and stability and their excessive cost.

Focusing on the first aspect, it will be mandatory to improve the performances of DSP systems by further optimization of their active materials. In terms of dyes, although a picture is starting to emerge regarding the need to precisely control their lipophilicity/hydrophilicity balance, as well as the precise position of their energy levels, more detailed design principles are required to develop structures able to work efficiently in the aqueous environment typical of DSP applications. Crucial aspects to be considered are the development of improved anchoring groups, able to ensure a rapid charge injection rate into the conduction band of the sensitizer, the use of panchromatic chromophores, to enhance light harvesting in the entirety of the visible spectrum, the attainment of a precise organization of dye molecules on the SC surface (also by the use of co-adsorbents), as well as the establishment of more efficient interactions with hole scavenger species present in solution, beyond the simple manipulation of orbital levels.

In parallel with the discovery of more efficient dye sensitizers, improvements are also required concerning the semiconductor structure. Significant results have already been obtained by exploring different $TiO_2$ polymorphs (either in pure form or as mixed phases) as well as precisely controlled titania nanostructures. Despite that, several problems still remain, such as the excessive rate of charge recombination favoured by the interaction of the hydrophilic $TiO_2$ surface with the aqueous reaction environment. They could be overcome by designing composites with improved dye/semiconductor/water interfaces, as well as by introducing semiconductors with tailored morphology to speed up charge transport and transfer to the HEC, thanks to the help of more refined kinetic models.[72]

Requirements of efficiency enhancement and cost reduction are closely linked to the need for shifting from model sacrificial electron donors, such as TEOA or EDTA, to more realistic species in terms of sustainability. As discussed above, this could be achieved by employing simple alcohols (*e. g.* EtOH, *i*PrOH) or biomass-derived reducing compounds (*e. g.* from glucose to more complex sugars, all the way to lignocellulose), either as intermediate solutions towards the ultimate goal of water splitting, or as platforms for coupling $H_2$ generation with production of value-added, oxidized compounds.

In terms of stability, a crucial aspect will be once again the development of improved anchoring groups, capable to ensure a robust attachment of the dye to the semiconductor surface with negligible hydrolysis, without an excessive limitation of performances, also by exploration of multi-branched structures. At the same time, it will also be necessary to design dyes devoid of labile functional groups, to avoid their oxidation or decomposition during the photocatalytic reaction.

Regarding cost reduction, a crucial step in this direction would be the replacement of platinum nanoparticles, usually employed as HEC, with cheaper and more readily available catalytic materials. Such modification would also eliminate the risk of shortages of catalytic material in the hypothesis of a future large-scale deployment of DSP technology. Although such work has already been done extensively in photoreforming studies using fully inorganic systems,[119] it has not yet been explored in depth in the field of DSP. Of particular interest is the work on supported molecular catalysts, as their structure can be tailored to make them very specific, opening the way to the development of parallel processes able to produce more than one compound at the same time: one such example is the dye-sensitized photocatalytic production of syngas ($H_2$+CO) recently published by Kang and co-workers.[120]

A summary of the most important recent advances in the fields and the possible directions of future development, as discussed in the above paragraphs, is provided in Figure 15.

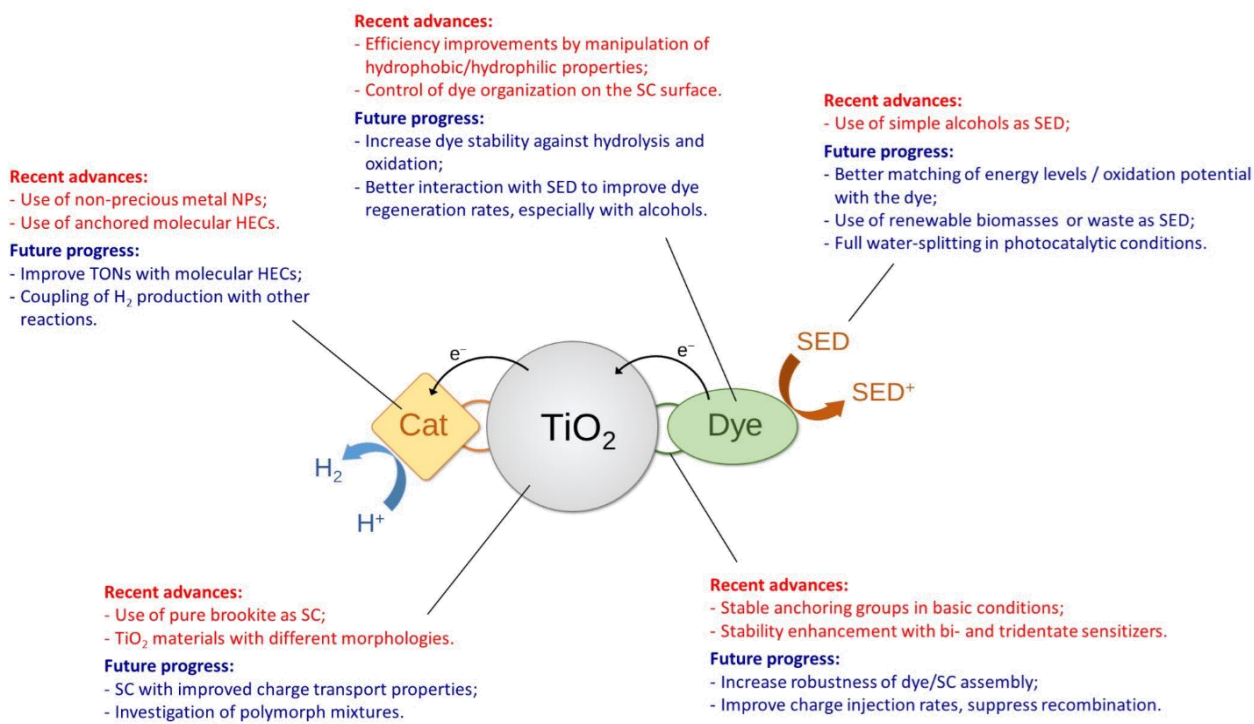

*Figure 15. Recent advances and potential future developments of DSP research.*

Finally, it is important that researchers working in the DSP field will try to adopt more consistent standards for experimental procedures, laboratory setups and data reporting. Despite significant efforts by publishers to promote "best practices" to perform measurements and data analysis, large discrepancies still remain in the way materials and devices properties are reported, sometimes preventing a meaningful comparison of results. It is mandatory to overcome these difficulties to ensure a correct future development of this research area.

As can be seen by the above discussion, despite the recent achievements documented in this article, many unsolved problems and open questions still wait to be addressed in the field of DSP $H_2$ production. With so much work to do, we have little doubt that it will remain a very active field of research for many years to come.

## 12. Acknowledgements

Financial support from European Community (Projects H2020 − RIA-CE-NMBP-25 Program − Grant No. 862030 and H2020-LC-SC3-2019-NZE-RES-CC − Grant No. 884444), University of Trieste, CNR-ICCOM ('SOLARSYNT' Project) and INSTM Consortium is acknowledged.

## 13. References

[1]     Balzani V, Credi A and Venturi M 2008 Photochemical Conversion of Solar Energy *ChemSusChem* **1** 26–58

[2]     Dau H, Fujita E and Sun L 2017 Artificial Photosynthesis: Beyond Mimicking Nature *ChemSusChem* **10** 4228–35

[3]     Armaroli N and Balzani V 2011 The Hydrogen Issue *ChemSusChem* **4** 21–36

[4]     Armaroli N and Balzani V 2016 Solar Electricity and Solar Fuels: Status and Perspectives in the Context of the Energy Transition *Chem. – A Eur. J.* **22** 32–57

[5]     Fujishima A and Honda K 1972 Electrochemical Photolysis of Water at a Semiconductor Electrode *Nature* **238** 37–8

[6]     Wang Z, Li C and Domen K 2019 Recent developments in heterogeneous photocatalysts for solar-driven overall water splitting *Chem. Soc. Rev.* **48** 2109–25

[7]     Carraro G, Maccato C, Gasparotto A, Montini T, Turner S, Lebedev O I, Gombac V, Adami G, Van Tendeloo G, Barreca D and Fornasiero P 2014 Enhanced Hydrogen Production by Photoreforming of Renewable Oxygenates Through Nanostructured Fe2O3 Polymorphs *Adv. Funct. Mater.* **24** 372–8

[8]     Uekert T, Kuehnel M F, Wakerley D W and Reisner E 2018 Plastic waste as a feedstock for solar-driven H2 generation *Energy Environ. Sci.* **11** 2853–7

[9]     Chen X, Liu L, Yu P Y and Mao S S 2011 Increasing solar absorption for photocatalysis with black hydrogenated titanium dioxide nanocrystals *Science (80-. ).* **331** 746–50

[10]    Maeda K 2013 Z - Scheme Water Splitting Using Two Di ff erent Semiconductor Photocatalysts **2**

[11]    Zhou P, Yu J and Jaroniec M 2014 All-solid-state Z-scheme photocatalytic systems *Adv. Mater.* **26** 4920–35

[12]    Zhang X, Peng T and Song S 2016 Recent advances in dye-sensitized semiconductor systems for photocatalytic hydrogen production *J. Mater. Chem. A* **4** 2365–402

[13]    Hagfeldt A, Boschloo G, Sun L, Kloo L and Pettersson H 2010 Dye-Sensitized Solar Cells *Chem. Rev.* **110** 6595–663

[14]    Li F, Yang H, Li W and Sun L 2018 Device Fabrication for Water Oxidation, Hydrogen Generation, and CO2 Reduction via Molecular Engineering *Joule* **2** 36–60

[15]    Li J and Wu N 2015 Semiconductor-based photocatalysts and photoelectrochemical cells for solar fuel generation: a review *Catal. Sci. Technol.* **5** 1360–84

[16]    Chen S, Takata T and Domen K 2017 Particulate photocatalysts for overall water splitting *Nat. Rev. Mater.* **2** 17050

[17]    Pellegrin Y and Odobel F 2017 Les donneurs d'électron sacrificiels pour la production de combustible solaire *Comptes Rendus Chim.* **20** 283–95

[18]    Watanabe M 2017 Dye-sensitized photocatalyst for effective water splitting catalyst *Sci. Technol. Adv. Mater.* **18** 705–23

[19]    Abe R, Shinmei K, Koumura N, Hara K and Ohtani B 2013 Visible-light-induced water splitting based on two-step photoexcitation between dye-sensitized layered niobate and tungsten oxide photocatalysts in the presence of a triiodide/iodide shuttle redox mediator *J. Am. Chem. Soc.* **135** 16872–84

[20]    Zhang X, Peng T, Yu L, Li R, Li Q and Li Z 2015 Visible/near-infrared-light-induced H2 production over g-C3N4 co-sensitized by organic dye and zinc phthalocyanine derivative *ACS Catal.* **5** 504–10

[21]    Yu L, Zhang X, Zhuang C, Lin L, Li R and Peng T 2014 Syntheses of asymmetric zinc

phthalocyanines as sensitizer of Pt-loaded graphitic carbon nitride for efficient visible/near-IR-light-driven H 2 production *Phys. Chem. Chem. Phys.* **16** 4106–14

[22] Ohtani B, Prieto-Mahaney O O, Li D and Abe R 2010 What is Degussa (Evonik) P25? Crystalline composition analysis, reconstruction from isolated pure particles and photocatalytic activity test *J. Photochem. Photobiol. A Chem.* **216** 179–82

[23] Qureshi M and Takanabe K 2017 Insights on measuring and reporting heterogeneous photocatalysis: Efficiency definitions and setup examples *Chem. Mater.* **29** 158–67

[24] Kunz L Y, Diroll B T, Wrasman C J, Riscoe A R, Majumdar A and Cargnello M 2019 Artificial inflation of apparent photocatalytic activity induced by catalyst-mass-normalization and a method to fairly compare heterojunction systems *Energy Environ. Sci.* **12** 1657–67

[25] Melchionna M and Fornasiero P 2020 Updates on the Roadmap for Photocatalysis *ACS Catal.* **10** 5493–501

[26] Melchionna M, Beltram A, Montini T, Monai M, Nasi L, Fornasiero P and Prato M 2016 Highly efficient hydrogen production through ethanol photoreforming by a carbon nanocone/Pd@TiO2 hybrid catalyst *Chem. Commun.* **52** 764–7

[27] Kozuch S and Martin J M L 2012 "Turning Over" Definitions in Catalytic Cycles *ACS Catal.* **2** 2787–94

[28] Kisch H and Bahnemann D 2015 Best Practice in Photocatalysis: Comparing Rates or Apparent Quantum Yields? *J. Phys. Chem. Lett.* **6** 1907–10

[29] Hoy J, Morrison P J, Steinberg L K, Buhro W E and Loomis R A 2013 Excitation Energy Dependence of the Photoluminescence Quantum Yields of Core and Core/Shell Quantum Dots *J. Phys. Chem. Lett.* **4** 2053–60

[30] Chatterjee D 2010 Effect of excited state redox properties of dye sensitizers on hydrogen production through photo-splitting of water over TiO2 photocatalyst *Catal. Commun.* **11** 336–9

[31] Suryani O, Higashino Y, Sato H and Kubo Y 2019 Visible-to-Near-Infrared Light-Driven Photocatalytic Hydrogen Production Using Dibenzo-BODIPY and Phenothiazine Conjugate as Organic Photosensitizer *ACS Appl. Energy Mater.* **2** 448–58

[32] Cecconi B, Manfredi N, Montini T, Fornasiero P and Abbotto A 2016 Dye-Sensitized Solar Hydrogen Production: The Emerging Role of Metal-Free Organic Sensitizers *European J. Org. Chem.* **2016** 5194–215

[33] Huang J-F, Lei Y, Luo T and Liu J-M 2020 Photocatalytic H2 Production from Water by Metal-free Dye-sensitized TiO2 Semiconductors: The Role and Development Process of Organic Sensitizers *ChemSusChem* **13** 5863–95

[34] Lee C P, Lin R Y Y, Lin L Y, Li C T, Chu T C, Sun S S, Lin J T and Ho K C 2015 Recent progress in organic sensitizers for dye-sensitized solar cells *RSC Adv.* **5** 23810–25

[35] Wu Y, Zhu W H, Zakeeruddin S M and Grätzel M 2015 Insight into D-A-π-A structured sensitizers: A promising route to highly efficient and stable dye-sensitized solar cells *ACS Appl. Mater. Interfaces*

    **7** 9307–18

[36]    Watanabe M, Hagiwara H, Iribe A, Ogata Y, Shiomi K, Staykov A, Ida S, Tanaka K and Ishihara T 2014 Spacer effects in metal-free organic dyes for visible-light-driven dye-sensitized photocatalytic hydrogen production *J. Mater. Chem. A* **2** 12952–61

[37]    Luo G G, Lu H, Wang Y H, Dong J, Zhao Y and Wu R B 2016 A D-π-A-π-A metal-free organic dye with improved efficiency for the application of solar energy conversion *Dye. Pigment.* **134** 498–505

[38]    Ho P Y, Mark M F, Wang Y, Yiu S C, Yu W H, Ho C L, McCamant D W, Eisenberg R and Huang S 2018 Panchromatic Sensitization with ZnII Porphyrin-Based Photosensitizers for Light-Driven Hydrogen Production *ChemSusChem* **11** 2517–28

[39]    Tiwari A, Krishna N V, Giribabu L and Pal U 2018 Hierarchical Porous TiO2 Embedded Unsymmetrical Zinc-Phthalocyanine Sensitizer for Visible-Light-Induced Photocatalytic H2 Production *J. Phys. Chem. C* **122** 495–502

[40]    Ho P-Y, Wang Y, Yiu S-C, Yu W-H, Ho C-L and Huang S 2017 Starburst Triarylamine Donor-Based Metal-Free Photosensitizers for Photocatalytic Hydrogen Production from Water *Org. Lett.* **19** 1048–51

[41]    Aslan E, Karaman M, Yanalak G, Can M, Ozel F and Patir I H 2019 The investigation of novel D-π-A type dyes (MK-3 and MK-4) for visible light driven photochemical hydrogen evolution *Dye. Pigment.* **171** 107710

[42]    Pastore M, Fantacci S and De Angelis F 2013 Modeling Excited States and Alignment of Energy Levels in Dye-Sensitized Solar Cells: Successes, Failures, and Challenges *J. Phys. Chem. C* **117** 3685–700

[43]    Zhang L and Cole J M 2015 Anchoring Groups for Dye-Sensitized Solar Cells *ACS Appl. Mater. Interfaces* **7** 3427–55

[44]    Warnan J, Willkomm J, Farré Y, Pellegrin Y, Boujtita M, Odobel F and Reisner E 2019 Solar electricity and fuel production with perylene monoimide dye-sensitised TiO 2 in water *Chem. Sci.* **10** 2758–66

[45]    Narayanaswamy K, Tiwari A, Mondal I, Pal U, Niveditha S, Bhanuprakash K and Singh S P 2015 Dithiafulvalene functionalized diketopyrrolopyrrole based sensitizers for efficient hydrogen production *Phys. Chem. Chem. Phys.* **17** 13710–8

[46]    Manzhos S, Jono R, Yamashita K, Fujisawa J, Nagata M and Segawa H 2011 Study of Interfacial Charge Transfer Bands and Electron Recombination in the Surface Complexes of TCNE, TCNQ, and TCNAQ with TiO2 *J. Phys. Chem. C* **115** 21487–93

[47]    Ooyama Y and Harima Y 2012 Photophysical and Electrochemical Properties, and Molecular Structures of Organic Dyes for Dye-Sensitized Solar Cells *ChemPhysChem* **13** 4032–80

[48]    Wang M, Grätzel C, Zakeeruddin S M and Grätzel M 2012 Recent developments in redox electrolytes for dye-sensitized solar cells *Energy Environ. Sci.* **5** 9394–405

[49]    Robson K C D, Hu K, Meyer G J and Berlinguette C P 2013 Atomic Level Resolution of Dye

Regeneration in the Dye-Sensitized Solar Cell *J. Am. Chem. Soc.* **135** 1961–71

[50]   Tiwari A, Duvva N, Rao V N, Venkatakrishnan S M, Giribabu L and Pal U 2019 Tetrathiafulvalene Scaffold-Based Sensitizer on Hierarchical Porous TiO 2 : Efficient Light-Harvesting Material for Hydrogen Production *J. Phys. Chem. C* **123** 70–81

[51]   Dessì A, Monai M, Bessi M, Montini T, Calamante M, Mordini A, Reginato G, Trono C, Fornasiero P and Zani L 2018 Towards Sustainable $H_2$Production: Rational Design of Hydrophobic Triphenylamine-based Dyes for Sensitized Ethanol Photoreforming *ChemSusChem* **11** 793–805

[52]   Lee J, Kwak J, Ko K C, Park J H, Ko J H, Park N, Kim E, Ryu D H, Ahn T K, Lee J Y and Son S U 2012 Phenothiazine-based organic dyes with two anchoring groups on TiO 2 for highly efficient visible light-induced water splitting *Chem. Commun.* **48** 11431–3

[53]   Manfredi N, Cecconi B, Calabrese V, Minotti A, Peri F, Ruffo R, Monai M, Romero-Ocaña I, Montini T, Fornasiero P and Abbotto A 2016 Dye-sensitized photocatalytic hydrogen production: Distinct activity in a glucose derivative of a phenothiazine dye *Chem. Commun.* **52** 6977–80

[54]   Manfredi N, Monai M, Montini T, Peri F, De Angelis F, Fornasiero P and Abbotto A 2018 Dye-Sensitized Photocatalytic Hydrogen Generation: Efficiency Enhancement by Organic Photosensitizer-Coadsorbent Intermolecular Interaction *ACS Energy Lett.* **3** 85–91

[55]   Martín C, Ziółek M and Douhal A 2016 Ultrafast and fast charge separation processes in real dye-sensitized solar cells *J. Photochem. Photobiol. C Photochem. Rev.* **26** 1–30

[56]   Watanabe M, Hagiwara H, Ogata Y, Staykov A, Bishop S R, Perry N H, Chang Y J, Ida S, Tanaka K and Ishihara T 2015 Impact of alkoxy chain length on carbazole-based, visible light-driven, dye sensitized photocatalytic hydrogen production *J. Mater. Chem. A* **3** 21713–21

[57]   Tiwari A, Mondal I and Pal U 2015 Visible light induced hydrogen production over thiophenothiazine-based dye sensitized TiO2 photocatalyst in neutral water *RSC Adv.* **5** 31415–21

[58]   Wang J, Chai Z, Liu S, Fang M, Chang K, Han M, Hong L, Han H, Li Q and Li Z 2018 Organic Dyes based on Tetraaryl-1,4-dihydropyrrolo-[3,2-b]pyrroles for Photovoltaic and Photocatalysis Applications with the Suppressed Electron Recombination *Chem. - A Eur. J.* **24** 18032–42

[59]   Huang J-F, Lei Y, Xiao L-M, Chen X-L, Zhong Y-H, Qin S and Liu J-M 2020 Photocatalysts for H2 Generation from Starburst Triphenylamine/Carbazole Donor-Based Metal-Free Dyes and Porous Anatase TiO2 Cube *ChemSusChem* **13** 1037–43

[60]   Bettucci O, Skaltsas T, Calamante M, Dessì A, Bartolini M, Sinicropi A, Filippi J, Reginato G, Mordini A, Fornasiero P and Zani L 2019 Combining Dithienosilole-Based Organic Dyes with a Brookite/Platinum Photocatalyst toward Enhanced Visible-Light-Driven Hydrogen Production *ACS Appl. Energy Mater.* **2** 5600–12

[61]   Lee S H, Park Y, Wee K R, Son H J, Cho D W, Pac C, Choi W and Kang S O 2010 Significance of hydrophilic characters of organic dyes in visible-light hydrogen generation based on TiO2 *Org. Lett.* **12** 460–3

[62]   Han W S, Wee K R, Kim H Y, Pac C, Nabetani Y, Yamamoto D, Shimada T, Inoue H, Choi H, Cho

K and Kang S O 2012 Hydrophilicity control of visible-light hydrogen evolution and dynamics of the charge-separated state in dye/TiO2/Pt hybrid systems *Chem. - A Eur. J.* **18** 15368–81

[63]  Bartolini M, Gombac V, Sinicropi A, Reginato G, Dessì A, Mordini A, Filippi J, Montini T, Calamante M, Fornasiero P and Zani L 2020 Tuning the Properties of Benzothiadiazole Dyes for Efficient Visible Light-Driven Photocatalytic H2 Production under Different Conditions *ACS Appl. Energy Mater.* **0**

[64]  Tiwari A and Pal U 2015 Effect of donor-donor-π-acceptor architecture of triphenylamine-based organic sensitizers over TiO2 photocatalysts for visible-light-driven hydrogen production *Int. J. Hydrogen Energy* **40** 9069–79

[65]  Peter L M and Upul Wijayantha K G 2014 Photoelectrochemical Water Splitting at Semiconductor Electrodes: Fundamental Problems and New Perspectives *ChemPhysChem* **15** 1983–95

[66]  Hisatomi T, Kubota J and Domen K 2014 Recent advances in semiconductors for photocatalytic and photoelectrochemical water splitting *Chem. Soc. Rev.* **43** 7520–35

[67]  Moss B, Lim K K, Beltram A, Moniz S, Tang J, Fornasiero P, Barnes P, Durrant J and Kafizas A 2017 Comparing photoelectrochemical water oxidation, recombination kinetics and charge trapping in the three polymorphs of TiO2 *Sci. Rep.* **7** 2938

[68]  Yamakata A, Vequizo J J M and Matsunaga H 2015 Distinctive Behavior of Photogenerated Electrons and Holes in Anatase and Rutile TiO2 Powders *J. Phys. Chem. C* **119** 24538–45

[69]  Günnemann C, Haisch C, Fleisch M, Schneider J, Emeline A V and Bahnemann D W 2019 Insights into Different Photocatalytic Oxidation Activities of Anatase, Brookite, and Rutile Single-Crystal Facets *ACS Catal.* **9** 1001–12

[70]  Beltram A, Romero-Ocaña I, Josè Delgado Jaen J, Montini T and Fornasiero P 2016 Photocatalytic valorization of ethanol and glycerol over TiO2 polymorphs for sustainable hydrogen production *Appl. Catal. A Gen.* **518** 167–75

[71]  Vequizo J J M, Matsunaga H, Ishiku T, Kamimura S, Ohno T and Yamakata A 2017 Trapping-Induced Enhancement of Photocatalytic Activity on Brookite TiO2 Powders: Comparison with Anatase and Rutile TiO2 Powders *ACS Catal.* **7** 2644–51

[72]  Sieland F, Schneider J and Bahnemann D W 2017 Fractal Charge Carrier Kinetics in TiO 2 *J. Phys. Chem. C* **121** 24282–91

[73]  Sieland F, Schneider J and Bahnemann D W 2018 Photocatalytic activity and charge carrier dynamics of TiO2 powders with a binary particle size distribution *Phys. Chem. Chem. Phys.* **20** 8119–32

[74]  Liu G, Yang H G, Pan J, Yang Y Q, Lu G Q (Max) and Cheng H-M 2014 Titanium Dioxide Crystals with Tailored Facets *Chem. Rev.* **114** 9559–612

[75]  Monai M, Montini T and Fornasiero P 2017 Brookite: Nothing new under the sun? *Catalysts* **7**

[76]  Di Paola A, Bellardita M and Palmisano L 2013 Brookite, the least known TiO2 photocatalyst *Catalysts* **3** 36–73

[77] Kusumawati Y, Hosni M, Martoprawiro M A, Cassaignon S and Pauporté T 2014 Charge Transport and Recombination in TiO2 Brookite-Based Photoelectrodes *J. Phys. Chem. C* **118** 23459–67

[78] Cargnello M, Montini T, Smolin S Y, Priebe J B, Jaén J J D, Doan-Nguyen V V T, McKay I S, Schwalbe J A, Pohl M M, Gordon T R, Lu Y, Baxter J B, Brückner A, Fornasiero P and Murray C B 2016 Engineering titania nanostructure to tune and improve its photocatalytic activity *Proc. Natl. Acad. Sci. U. S. A.* **113** 3966–71

[79] Xie Y P, Yu Z B, Liu G, Ma X L and Cheng H-M 2014 CdS–mesoporous ZnS core–shell particles for efficient and stable photocatalytic hydrogen evolution under visible light *Energy Environ. Sci.* **7** 1895–901

[80] Bae E, Choi W, Park J, Shin H S, Kim S Bin and Lee J S 2004 Effects of Surface Anchoring Groups (Carboxylate vs Phosphonate) in Ruthenium-Complex-Sensitized TiO2 on Visible Light Reactivity in Aqueous Suspensions *J. Phys. Chem. B* **108** 14093–101

[81] Bae E and Choi W 2006 Effect of the Anchoring Group (Carboxylate vs Phosphonate) in Ru-Complex-Sensitized TiO2 on Hydrogen Production under Visible Light *J. Phys. Chem. B* **110** 14792–9

[82] Warnan J, Willkomm J, Ng J N, Godin R, Prantl S, Durrant J R and Reisner E 2017 Solar H2 evolution in water with modified diketopyrrolopyrrole dyes immobilised on molecular Co and Ni catalyst-TiO2 hybrids *Chem. Sci.* **8** 3070–9

[83] Abe R, Hara K, Sayama K, Domen K and Arakawa H 2000 Steady hydrogen evolution from water on Eosin Y-fixed TiO2 photocatalyst using a silane-coupling reagent under visible light irradiation *J. Photochem. Photobiol. A Chem.* **137** 63–9

[84] Brennan B J, Llansola Portolés M J, Liddell P A, Moore T A, Moore A L and Gust D 2013 Comparison of silatrane, phosphonic acid, and carboxylic acid functional groups for attachment of porphyrin sensitizers to TiO2 in photoelectrochemical cells *Phys. Chem. Chem. Phys.* **15** 16605–14

[85] Castellucci E, Monini M, Bessi M, Iagatti A, Bussotti L, Sinicropi A, Calamante M, Zani L, Basosi R, Reginato G, Mordini A, Foggi P and Di Donato M 2017 Photoinduced excitation and charge transfer processes of organic dyes with siloxane anchoring groups: A combined spectroscopic and computational study *Phys. Chem. Chem. Phys.* **19**

[86] Kakiage K, Aoyama Y, Yano T, Oya K, Fujisawa J and Hanaya M 2015 Highly-efficient dye-sensitized solar cells with collaborative sensitization by silyl-anchor and carboxy-anchor dyes *Chem. Commun.* **51** 15894–7

[87] Sobuś J, Gierczyk B, Burdziński G, Jancelewicz M, Polanski E, Hagfeldt A and Ziółek M 2016 Factors Affecting the Performance of Champion Silyl-Anchor Carbazole Dye Revealed in the Femtosecond to Second Studies of Complete ADEKA-1 Sensitized Solar Cells *Chem. – A Eur. J.* **22** 15807–18

[88] Choi S K, Yang H S, Kim J H and Park H 2012 Organic dye-sensitized TiO 2 as a versatile photocatalyst for solar hydrogen and environmental remediation *Appl. Catal. B Environ.* **121–122**

206–13

[89]   Watanabe M, Sun S, Ishihara T, Kamimura T, Nishimura M and Tani F 2018 Visible Light-Driven Dye-Sensitized Photocatalytic Hydrogen Production by Porphyrin and its Cyclic Dimer and Trimer: Effect of Multi-Pyridyl-Anchoring Groups on Photocatalytic Activity and Stability *ACS Appl. Energy Mater.* **1** 6072–81

[90]   Ding H, Xu M, Zhang S, Yu F, Kong K, Shen Z and Hua J 2020 Organic blue-colored D-A-π-A dye-sensitized TiO2 for efficient and stable photocatalytic hydrogen evolution under visible/near-infrared-light irradiation *Renew. Energy* **155** 1051–9

[91]   Cecconi B, Manfredi N, Ruffo R, Montini T, Romero-Ocaña I, Fornasiero P and Abbotto A 2015 Tuning Thiophene-Based Phenothiazines for Stable Photocatalytic Hydrogen Production *ChemSusChem* **8** 4216–28

[92]   Li Q, Che Y, Ji H, Chen C, Zhu H, Ma W and Zhao J 2014 Ortho-Dihydroxyl-9,10-anthraquinone dyes as visible-light sensitizers that exhibit a high turnover number for hydrogen evolution *Phys. Chem. Chem. Phys.* **16** 6550–4

[93]   Lai H, Liu X, Zeng F, Peng G, Li J and Yi Z 2020 Multicarbazole-Based D−π–A Dyes Sensitized Hydrogen Evolution under Visible Light Irradiation *ACS Omega* **5** 2027–33

[94]   Huang J-F, Liu J-M, Xiao L-M, Zhong Y-H, Liu L, Qin S, Guo J and Su C-Y 2019 Facile synthesis of porous hybrid materials based on Calix-3 dye and TiO2 for high photocatalytic water splitting performance with excellent stability *J. Mater. Chem. A* **7** 2993–9

[95]   Zhang X, Veikko U, Mao J, Cai P and Peng T 2012 Visible-Light-Induced Photocatalytic Hydrogen Production over Binuclear RuII–Bipyridyl Dye-Sensitized TiO2 without Noble Metal Loading *Chem. – A Eur. J.* **18** 12103–11

[96]   Kruth A, Hansen S, Beweries T, Brüser V and Weltmann K-D 2013 Plasma Synthesis of Polymer-Capped Dye-Sensitised Anatase Nanopowders for Visible-Light-Driven Hydrogen Evolution *ChemSusChem* **6** 152–9

[97]   Puga A V, Forneli A, García H and Corma A 2014 Production of H2 by Ethanol Photoreforming on Au/TiO2 *Adv. Funct. Mater.* **24** 241–8

[98]   Imizcoz M and Puga A V 2019 Optimising hydrogen production via solar acetic acid photoreforming on Cu/TiO2 *Catal. Sci. Technol.* **9** 1098–102

[99]   Imizcoz M and Puga A V 2019 Assessment of Photocatalytic Hydrogen Production from Biomass or Wastewaters Depending on the Metal Co-Catalyst and Its Deposition Method on TiO2 *Catalysts* **9** 584

[100]   Jin Z, Zhang X, Li Y, Li S and Lu G 2007 5.1% Apparent quantum efficiency for stable hydrogen generation over eosin-sensitized CuO/TiO2 photocatalyst under visible light irradiation *Catal. Commun.* **8** 1267–73

[101]   Le T T, Akhtar M S, Park D M, Lee J C and Yang O-B 2012 Water splitting on Rhodamine-B dye sensitized Co-doped TiO2 catalyst under visible light *Appl. Catal. B Environ.* **111**–**112** 397–401

[102] Yan Z, Yu X, Zhang Y, Jia H, Sun Z and Du P 2014 Enhanced visible light-driven hydrogen production from water by a noble-metal-free system containing organic dye-sensitized titanium dioxide loaded with nickel hydroxide as the cocatalyst *Appl. Catal. B Environ.* **160**–**161** 173–8

[103] Aslan E, Gonce M K, Yigit M Z, Sarilmaz A, Stathatos E, Ozel F, Can M and Patir I H 2017 Photocatalytic H2 evolution with a Cu2WS4 catalyst on a metal free D-π-A organic dye-sensitized TiO2 *Appl. Catal. B Environ.* **210** 320–7

[104] Patir I H, Aslan E, Yanalak G, Karaman M, Sarilmaz A, Can M, Can M and Ozel F 2019 Donor-Π-acceptor dye-sensitized photoelectrochemical and photocatalytic hydrogen evolution by using Cu 2 WS 4 co-catalyst *Int. J. Hydrogen Energy* **44** 1441–50

[105] Aslan E, Karaman M, Yanalak G, Bilgili H, Can M, Ozel F and Patir I H 2020 Synthesis of novel tetrazine based D-π-A organic dyes for photoelectrochemical and photocatalytic hydrogen evolution *J. Photochem. Photobiol. A Chem.* **390** 112301

[106] Lakadamyali F and Reisner E 2011 Photocatalytic H2 evolution from neutral water with a molecular cobalt catalyst on a dye-sensitised TiO2 nanoparticle *Chem. Commun.* **47** 1695–7

[107] Lakadamyali F, Reynal A, Kato M, Durrant J R and Reisner E 2012 Electron Transfer in Dye-Sensitised Semiconductors Modified with Molecular Cobalt Catalysts: Photoreduction of Aqueous Protons *Chem. – A Eur. J.* **18** 15464–75

[108] Willkomm J, Muresan N M and Reisner E 2015 Enhancing H2 evolution performance of an immobilised cobalt catalyst by rational ligand design *Chem. Sci.* **6** 2727–36

[109] Gross M A, Reynal A, Durrant J R and Reisner E 2014 Versatile Photocatalytic Systems for H2 Generation in Water Based on an Efficient DuBois-Type Nickel Catalyst *J. Am. Chem. Soc.* **136** 356–66

[110] Kudo A and Miseki Y 2009 Heterogeneous photocatalyst materials for water splitting *Chem. Soc. Rev.* **38** 253–78

[111] Lhermitte C R and Sivula K 2019 Alternative Oxidation Reactions for Solar-Driven Fuel Production *ACS Catal.* **9** 2007–17

[112] Chen X, Shen S, Guo L and Mao S S 2010 Semiconductor-based Photocatalytic Hydrogen Generation *Chem. Rev.* **110** 6503–70

[113] Beltram A, Melchionna M, Montini T, Nasi L, Fornasiero P and Prato M 2017 Making H2 from light and biomass-derived alcohols: the outstanding activity of newly designed hierarchical MWCNT/Pd@TiO2 hybrid catalysts *Green Chem.* **19** 2379–89

[114] Li X, Cui S, Wang D, Zhou Y, Zhou H, Hu Y, Liu J G, Long Y, Wu W, Hua J and Tian H 2014 New organic donor-acceptor-π-acceptor sensitizers for efficient dye-sensitized solar cells and Photocatalytic hydrogen evolution under visible-light irradiation *ChemSusChem* **7** 2879–88

[115] Maitani M M, Zhan C, Mochizuki D, Suzuki E and Wada Y 2013 Influence of co-existing alcohol on charge transfer of H2 evolution under visible light with dye-sensitized nanocrystalline TiO2 *Appl. Catal. B Environ.* **140**–**141** 406–11

[116]   Cook A W and Waldie K M 2019 Molecular Electrocatalysts for Alcohol Oxidation: Insights and Challenges for Catalyst Design *ACS Appl. Energy Mater.*

[117]   Pho T V., Sheridan M V., Morseth Z A, Sherman B D, Meyer T J, Papanikolas J M, Schanze K S and Reynolds J R 2016 Efficient Light-Driven Oxidation of Alcohols Using an Organic Chromophore-Catalyst Assembly Anchored to TiO2 *ACS Appl. Mater. Interfaces* **8** 9125–33

[118]   Kuehnel M F and Reisner E 2018 Solar Hydrogen Generation from Lignocellulose *Angew. Chemie - Int. Ed.* **57** 3290–6

[119]   Christoforidis K C and Fornasiero P 2017 Photocatalytic Hydrogen Production: A Rift into the Future Energy Supply *ChemCatChem* **9** 1523–44

[120]   Lee J S, Won D Il, Jung W J, Son H J, Pac C and Kang S O 2017 Widely Controllable Syngas Production by a Dye-Sensitized TiO2Hybrid System with ReIand CoIIICatalysts under Visible-Light Irradiation *Angew. Chemie - Int. Ed.* **56** 976–80